\documentclass[sigconf]{acmart}

\usepackage{microtype}
\usepackage{graphicx}
\usepackage{subfigure}
\usepackage{booktabs} %
\usepackage{setspace}
\usepackage{xcolor}
\usepackage{multirow}
\usepackage{adjustbox}
\usepackage{multicol}
\usepackage{booktabs}
\usepackage{threeparttable}
\usepackage{tabulary}
\usepackage{colortbl}
\usepackage{ragged2e}
\usepackage{enumitem}
\usepackage{placeins}
\usepackage{afterpage}
\usepackage{float}
\usepackage{dsfont}
\usepackage{amsmath,bm}
\usepackage{caption} %

\newcommand{\method}{PAAG}
\newcommand{\dataset}{ProtAnnotation}
\newcommand{\eat}[1]{}
\theoremstyle{definition}
\newtheorem{definition}{Definition}[section]
\AtBeginDocument{%
  \providecommand\BibTeX{{%
    \normalfont B\kern-0.5em{\scshape i\kern-0.25em b}\kern-0.8em\TeX}}}

\setcopyright{acmlicensed}
\copyrightyear{2018}
\acmYear{2018}
\acmDOI{XXXXXXX.XXXXXXX}

\acmConference[Conference acronym 'XX]{Make sure to enter the correct
  conference title from your rights confirmation emai}{June 03--05,
  2018}{Woodstock, NY}
\acmISBN{978-1-4503-XXXX-X/18/06}

\begin{document}

\title{Annotation-guided Protein Design with Multi-Level Domain Alignment}

\author{Chaohao Yuan}
\authornote{Work was done when Chaohao Yuan worked as an intern at Tencent AI Lab.}
\affiliation{%
  \institution{Tsinghua University}
    \city{Shenzhen}
  \country{China}
}
\email{yuanch22@mails.tsinghua.edu.cn}

\author{Songyou Li}
\affiliation{%
  \institution{Renmin University of China}
  \city{Beijing}
  \country{China}
}
\email{songyou_li@126.com}

\author{Geyan Ye}
\authornote{Project Lead.}
\affiliation{%
  \institution{Tencent AI Lab}
  \city{Shenzhen}
  \country{China}
}
\email{blazerye@tencent.com}

\author{Yikun Zhang}
\affiliation{%
 \institution{Peking University}
   \city{Shenzhen}
  \country{China}
 }
\email{yikun.zh@hotmail.com}

\author{Long-Kai Huang}
\affiliation{%
  \institution{Tencent AI Lab}
  \city{Shenzhen}
  \country{China}
  }
\email{hlongkai@gmail.com}

\author{Wenbing Huang}
\affiliation{%
  \institution{Renmin University of China}
  \city{Beijing}
  \country{China}
  }
\email{hwenbing@126.com}

\author{Wei Liu}
\affiliation{%
  \institution{Tencent AI Lab}
  \city{Shenzhen}
  \country{China}
  }
\email{topliu@tencent.com}

\author{Jianhua Yao}
\affiliation{%
  \institution{Tencent AI Lab}
  \city{Shenzhen}
  \country{China}
  }
\email{jianhua.yao@gmail.com}

\author{Yu Rong}
\authornote{Corresponding Author.}
\affiliation{%
  \institution{DAMO Academy, Alibaba Group}
  \city{Hangzhou}
  \country{China}
  }
\email{yu.rong@hotmail.com}

\renewcommand{\shortauthors}{Trovato and Tobin, et al.}

\begin{abstract}
The core challenge of \emph{de novo} protein design lies in creating proteins with specific functions or properties, guided by certain conditions. Current models explore to generate protein using structural and evolutionary guidance, which only provide indirect conditions concerning functions and properties. However, textual annotations of proteins, especially the annotations for protein domains, which directly describe the protein's high-level functionalities, properties, and their correlation with target amino acid sequences, remain unexplored in the context of protein design tasks.
In this paper, we propose \textbf{P}rotein-\textbf{A}nnotation \textbf{A}lignment \textbf{G}eneration (\method), a multi-modality protein design framework that integrates the textual annotations extracted from protein database for controllable generation in sequence space. Specifically, within a multi-level alignment module, {\method} can explicitly generate proteins containing specific domains conditioned on the corresponding domain annotations, and can even design novel proteins with flexible combinations of different kinds of annotations. Our experimental results underscore the superiority of the aligned protein representations from {\method} over 7 prediction tasks. Furthermore, {\method} demonstrates a significant increase in generation
success rate (24.7\% vs 4.7\% in zinc finger, and 54.3\% vs 22.0\% in the immunoglobulin domain) in comparison to the existing model.
We anticipate that {\method} will broaden the horizons of protein design by leveraging the knowledge from between textual annotation and proteins.
\end{abstract}



\maketitle

\section{Introduction}
Protein design~\cite{marshall2019minimalist} is a crucial task for its immense potential on drug discovery~\cite{sliwoski2014computational,wu2024fast}, enzyme engineering~\cite{STERNER2008421}, immunongineering~\cite{swartz2012engineering} and so on. The generation of proteins with specific properties, behaviors, or functions, such as optimizing the binding affinity to given molecules~\cite{houk2003binding,zhao2023geometric} or incorporating a particular ion-binding site~\cite{regan1995protein}, is known as \emph{de novo} protein design. This process presents a significant challenge due to the vast space of protein sequences and the complexity of protein functions. Recently, machine learning models have shown profound potential for protein design. The existing studies mostly rely on the structural~\cite{watson2023novo} or evolutionary information~\cite{alamdari2023protein} as the guidance to design proteins. 
However, in many cases, these conditions can only offer indirect guidance towards the desired protein design targets to their inherent ambiguity. For example, the same protein sequence segment
can be either act as receptors to regulate synaptic function~\cite{stathakis1997human} or helpers to locate target proteins to specific subcellular locations~\cite{MOHANDAS1995594}.

\begin{figure}[t]
\begin{center}
\centerline{\includegraphics[width=0.90\linewidth]{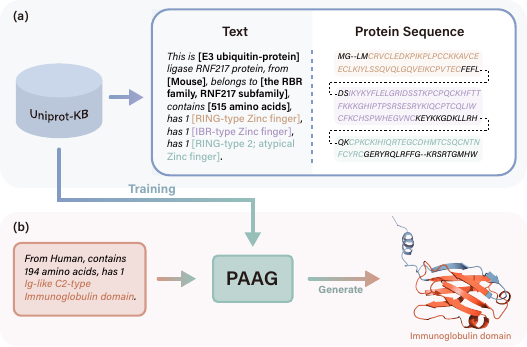}}
\vspace{-2ex}
\caption{ (a) The example of property annotations (in bold) and domain annotations (in colors). (b). The illustration of annotation-guided protein design with {\method}. Given the input of textual description within immunoglobulin domain annotation, {\method} can generate the proteins containing immunoglobulin domain.
}
\label{fig:intro_overview}
\end{center}
\vspace{-6ex}
\end{figure}

In addition to the structural and evolutionary information, the current protein dataset, such as Swiss-Prot~\cite{bairoch2000swiss} and UniProtKB~\cite{10.1093/nar/gkac1052}, contains rich textual annotations derived from wet laboratory experiments and literature. Figure~\ref{fig:intro_overview} illustrates an example of the textual annotations on the zinc-finger protein. Generally, these annotations can be categorized into \emph{property annotation} and \emph{domain annotation}. \emph{Property annotation} represents a piece of text that depicts the global property of proteins, such as protein names, number of amino acids, subcellular localization~\cite{almagro2017deeploc} and thermostability~\cite{dallago2021flip}. Conversely, \emph{domain annotation} pertains to the knowledge derived from the local domain~\cite{drenth1968structure} of proteins, which is a subregion of amino acid sequence that is self-stabilizing and represents certain structural and functional aspects of the protein.  These annotations provide both coarse and fine-grained information regarding protein's functions, properties, and interactions, thereby encompassing knowledge with the potential to guide the generation and design of novel proteins. 

For instance, as one of crucial functional domains of DNA/RNA binding proteins~\cite{cassandri2017zinc}, the zinc-finger domain naturally has many variants, such as C2H2 type, CCHC type and Zinc ribbon type. These variants 
exhibit significant differences in both structural and evolutionary features among them which is hard covered by structural and evolutionary conditions. On the contrary, the ``zinc-finger'' annotation from the protein database inherently provides a more effective means of describing the high-level knowledge span across both sequences and structures. Hence, we aim to investigate the following question: \emph{Is it possible to leverage such textual annotations to guide the delicate controllable protein design?}

Recently, several primary attempts have been made to leverage such textual annotations to guide the protein generation. 
Examples include
training an individual protein caption model to guide the diffusion generation process~\cite{ingraham2022chroma}, and incorporating the an overall text description through a global language-to-protein alignment model~\cite{liu2023proteindt}. 
However, current models cannot flexibly combine the different conditions and lack of the capability for fine-grained control, such as specifying the generation of particular domains.

To fill these gaps, in this paper, we introduce a novel framework, \textbf{P}rotein-\textbf{A}nnotation \textbf{A}lignment \textbf{G}eneration(\method), that enables annotation-guided protein design by aligning protein sequences with their textual annotations. Specifically, we first consider both property and domain annotations in proteins and design a multi-level alignment module to align the representations of sequences and annotations extracted from the existing encoders in both global and local level. For the generation task, {\method} utilizes an autoregressive decoder to generate protein sequences guided by the aligned representation of textual annotations. Additionally, {\method} employs an end-to-end training pipeline that joint training of alignment and generation tasks without freezing the parameters in sequence and text encoders. This joint training enhances the understanding of the complex and flexible annotation condition, resulting in improved guided generation. Figure~\ref{fig:intro_overview} demonstrates an example of annotation-guided generation. 
In experiments, we first investigate the quality of protein representations from {\method} by seven predictive downstream tasks. {\method} surpasses state-of-the-art baseline with an average relative improvement of 1.5\%.  Subsequently, three protein generation tasks are conducted to assess the capabilities of {\method}. In the case of unconditional generation, {\method} produces sequences exhibiting the highest degree of novelty while maintaining the distribution of natural proteins. For the two conditional protein design tasks, {\method} demonstrates a nearly threefold increase in generation success rate (24.7\% vs 4.7\% in zinc finger, and 54.3\% vs 22.0\% in the immunoglobulin domain)\footnote{This is the success rate with quality threshold $e=1$, i.e., $\text{SR}_{1}$.} in comparison to the existing model.
Our contributions are summarized as follows:
\begin{itemize}[leftmargin=*]
    \item We propose the first annotation-guided protein design paradigm, integrating local and global annotation information. Our proposed framework {\method} is the first approach that can generate proteins containing specific domains, guided by their corresponding annotations with high success rate.
    \item  {\method} features a multi-level alignment module for handling annotation and protein data alignment at various granularities. Joint training of alignment and generation tasks allows the model to produce improved protein representations, consequently boosting performance in predictive and generative tasks.
    
    \item Comprehensive experiments on 7 predictive and 3 generative tasks showcase {\method}'s superiority compared to the existing methods. Notably, {\method} is not only capable of generating proteins that include a single annotation, but it also successfully generates proteins adhering to the flexible conditions of multiple annotation combinations.
\end{itemize}

\section{Preliminaries}

\subsection{Protein and Its Textual Annotations} 
The primary structure of a protein can be represented as an amino acid sequence $S=(s_1, s_2, \cdots, s_n)$, where $s_i$ is the $i$-th amino acids chosen from 20 different amino acids which are represented as 20 characters. 
Given a protein $S$, the annotation-sequence pair set $\mathcal{D}_{S}=\{(T_k, S^k_{i:j})\}_{k=1}^{K_S}$ is constructed by extracting the correspondence between textual annotations and protein domains \& properties from protein database, such as UniProtKB~\cite{10.1093/nar/gkac1052}, where $T_k$ is the textual annotations of the $k$-th domain $S^k_{i:j}$ of the protein and $K_S$ is the number of annotation-domain pairs of the protein $S$. For the property annotation,  the corresponding domain is the entire sequence, i.e, $S_{1: |S|}$.  This annotation-domain pair set $\mathcal{D}_{S}$ provides comprehensive knowledge about the protein $S$ and, therefore, plays an important role in training the protein generation model.

\subsection{Encoders for Proteins and Annotations}
\label{Protein and Annotation Encoder}

We adopt a protein encoder (PE) to generate both the protein and its domain representations based on the amino acid sequence as $\bm{z}_{S} = f_{\text{PE}}(S)$, where $f_{\text{PE}}$ is the protein encoder and $\bm{z}_{S}$ is the embedding of $S$.  
For textual annotations, we generate their representations using the pre-trained language model $f_{\text{LM}}$ as $\bm{z}_{\mathcal{A}_S} = f_{\text{LM}}(G(\mathcal{A}_S))$, where $\mathcal{A}_S \subseteq \{T_k | (T_k, S_{i:j}^k) \in \mathcal{D}_S \}$ is a subset of the annotations in the annotation-domain pair set for a protein $S$, $G()$ is a template function which converts a set of annotations to a textual description, the details of $G()$ can be found in Appendix \ref{appendix.dataset}, and $\bm{z}_{\mathcal{A}}$ is the embedding of the annotation set. Note that the annotation subset can cover all annotations in $\mathcal{D}_S$ or just one. If $\mathcal{A}_S$ covers only one annotation $T_0$, we can skip the template function and directly generate the embedding  of $T_0$ using $f_{\text{LM}}$.

We employ transformer-based protein encoder and text encoder as our base model and initialize them with pre-trained models. Specifically, we employ SciBERT\cite{beltagy2019scibert}, which is pre-trained on computer science and biological datasets, to initialize the text encoder $f_{\text{LM}}$. The pre-trained protein encoder, ProtBERT\cite{elnaggar2021protrans}, is used to initialize the protein encoder $f_{\text{PE}}$.

\begin{figure*}[h]
\begin{center}
\centerline{\includegraphics[width=0.82
\linewidth]{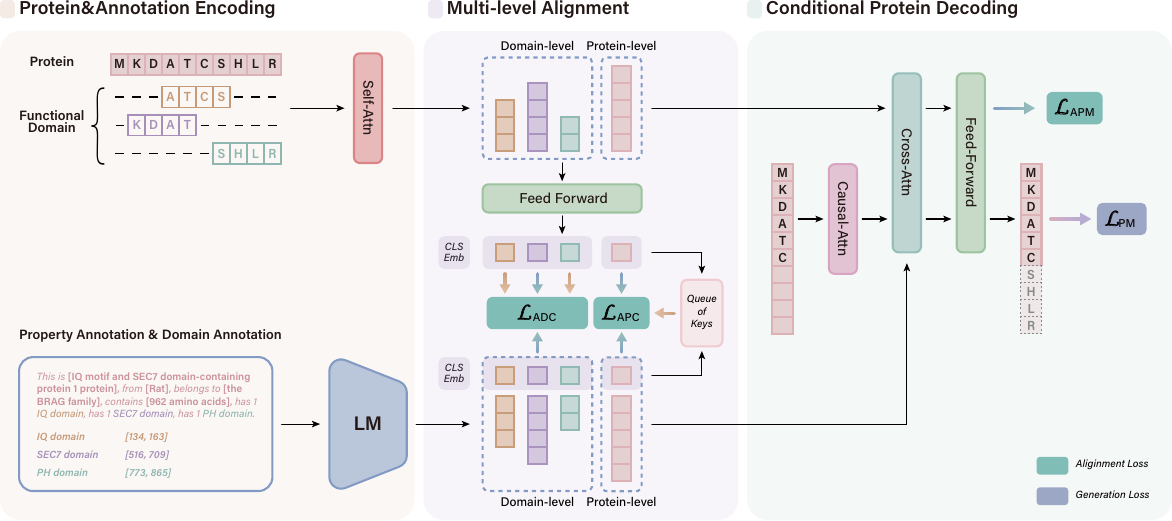}}
\vspace{-2ex}
\caption{The overall framework of {\method}. The same parameters share the same color. {\method} contains three modules. (1) Protein \& Annotation Encoding module encode the input protein sequence \& domains and corresponding annotations to the embeddings. (2) Multi-level alignment module projects the protein and annotation embeddings into  and employs Annotation-Protein Contrasive (APC) loss, Annotation-Domain Contrasive (ADC) loss and Annotation-Protein Matching (APM) loss to align them in a same latent space. (3) Conditional Protein Decoding accepts the annotation embedding as input and generate the protein sequence. }
\label{fig:framework}
\end{center}
\vspace{-6ex}
\end{figure*}

\section{Methodology}
\label{Methodology}
In this section, we propose a novel framework, \textbf{P}rotein-\textbf{A}nnotation \textbf{A}lignment \textbf{G}eneration(\method), which enables the flexible annotations-guided protein design. Figure~\ref{fig:framework} illustrates the overall framework of {\method}. 
In the following, we present the details of each component of {\method}.

\subsection{Multi-level Protein and Annotation Alignment}
\label{Multi-level Protein and Annotation Alignment}
We will first employ template function $G$ to translate these annotations as textual descriptions and then utilize a language model to extract the representation of the annotation set $\bm{z}_{\mathcal{A}} = f_{\text{LM}}(G(\mathcal{A}))$. Then the protein sequence is generated using a decoder $f_{D}$ based on this representation as $S = f_{\text{D}}(\bm{z}_{\mathcal{A}})$.

To better integrate the information between proteins and annotations, we aim to align the multi-level representations of proteins and annotations. Specifically, we conduct local alignment and global alignment by performing contrastive learning at domain level and protein level, respectively.

We measure the alignment score using the cosine similarity between the embeddings of protein and the annotation set, which is defined as 
\vspace{-1ex}
\begin{align}
\vspace{-1ex}
    s(\mathcal{A}, S) = \frac{\langle \text{Proj}_{\mathcal{A}}(z_{\mathcal{A}}), \text{Proj}_{S}(\bm{z}_S) \rangle}{\| \text{Proj}_{\mathcal{A}}(\bm{z}_{\mathcal{A}}) \| \|\text{Proj}_{S}(\bm{z}_S) \|},
\end{align}
where $\text{Proj}_{a}(\bm{z}) = \bm{W}_{a}\bm{z} + \bm{b}$ projects the input $\bm{z}$ into a latent space with dimension $h$. $\bm{W}_{a}\in \mathbb{R}^{h \times |\bm{z}|}$ and $\bm{b} \in \mathbb{R}^{h}$ are trainable parameters. 
During alignment, we encourage the matched (positive) pairs to have representations with higher similarity $s$ than unmatched (negative) pairs. We next will explain in detail how to construct the positive and negative pairs in our multi-level framework.

\subsubsection{Local Alignment} 

Given a protein $S$, and its annotation-domain pairs set $\mathcal{D}_{S}$, in domain-level, we aim to align the representation of the domain $S^k_{i:j}$ and its corresponding annotation $T_k$ and use all annotation-domain pairs in $\mathcal{D}_{S}$ as positive pair.
To construct the negative pairs, we randomly sample the sub-regions outside the domain $S^k_{i:j}$, i.e. $S^k_{1:i-1} \cup S^k_{j+1:l}$, as the negative samples $S^{k-}_{i:j}$ of the domain. 

\textbf{Annotation-Domain Contrastive(ADC) Loss:} 
In designing this loss, since our objective is on identifying functional regions within proteins, we consider amino acid sequences from the same protein as negative samples. Specifically, we adopt InfoNCE loss as the ADC loss to align the representations of local annotation $\bm{z}_T$ and functional domain $\bm{z}_{S_{i:j}}$ as
\vspace{-1ex}
\begin{equation} \label{eqn:infonce}
\vspace{-1ex}
\begin{split}
\mathcal{L}_{\mathrm{ADC}} = - \frac{1}{K_S} \sum_{k=1}^{K_S} %
& \log \frac{\exp(s({T_k}, {S^k_{i:j}}) / \tau)}{\sum_{n=1}^N \exp(s({T_n}, {S^{n-}_{i:j}}) / \tau)}%
\end{split},
\end{equation}
where $K_S$ is the number of functional domains for the protein $S$, $N$ is the number of negative samples, and $\tau$ is a learnable temperature parameter.

The local alignment enables {\method} to explicitly learn the relation between the functional domain and its annotation. Therefore, we can use the model to controllably generate the specific functional domain given annotations.

\subsubsection{Global Alignment} 

To enable global alignment, we utilize template function $G(\cdot)$ to construct protein-level textual description and form the positive protein-description pairs as $\{(G(\mathcal{A}_S), S)\}$. %
Since multiple annotations and the entire protein will be more complex, to enlarge the number of negative samples for global alignment, we follow the setting of MOCO \cite{he2020momentum} to construct momentum encoders $f_{m}$ as:
\vspace{-1ex}
\begin{equation}
\label{momentum}
f_m \leftarrow m f_m + (1 - m) f.
\end{equation}
where the encoder $f$ can be either protein encoder $f_{\text{PE}}$ or text encoder $f_{\text{LM}}$, and $m$ is the momentum hyperparameter. We follow implementation details in \cite{li2021align} and \cite{li2022blip} to construct the momentum encoders for both encoders.
The momentum encoders extract consistent features to increase the number of negative samples, and the dynamic dictionaries will store these features.  %

\textbf{Annotation-Protein Contrastive (APC) Loss} is designed to align the representations of global properties $\bm{z}_{\mathcal{A}}$ with protein $\bm{z}_S$. %
Specifically, for each protein sequence and annotation set, we calculate the softmax-normalized sequence-to-annotation and annotation-to-sequence similarity as:
\vspace{-1ex}
\begin{equation}
\vspace{-1ex}
\label{eqn:sim}
p_m^\mathrm{s2a}(S) = \frac{\exp (s(A_m,S) / \tau)}{\sum_{m=1}^M \exp (s(A_m,S)/ \tau)},
\end{equation}
\vspace{-0.5ex}
\begin{equation}
\vspace{-0.5ex}
\label{eqn:sim}
p_m^\mathrm{a2s}(A) = \frac{\exp (s(A,S_m)/ \tau)}{\sum_{m=1}^M \exp (s(A,S_m)/ \tau)},
\end{equation}
where $\tau$ is a learnable temperature parameter, $A_m$ and $S_m$ indicate their representations will be extracted by respective momentum encoders. 

Denote $\Vec{y}^\mathrm{~a2s}(A)$ and $\Vec{y}^\mathrm{~s2a}(S)$ as the ground-truth one-hot similarity,
where negative pairs have a probability of 0 and the positive pair has a probability of 1.
The annotation-protein contrastive loss is defined as the cross-entropy $\mathrm{H}$ between $\Vec{p}$ and $\Vec{y}$:
\vspace{-1ex}
\begin{equation}
\vspace{-1ex}
\label{eqn:APC}
\begin{split}
\mathcal{L}_\mathrm{APC} = \frac{1}{2} \mathbb{E}_{(S,A)\sim D} \big[ \mathrm{H}(\Vec{y}^\mathrm{~s2a}(S),\Vec{p}^\mathrm{~s2a}(S)) \\+ \mathrm{H}(\Vec{y}^\mathrm{~a2s}(A),\Vec{p}^\mathrm{~a2s}(A)) \big]
\end{split}.
\end{equation}

Through APC loss, PAAG aligns the representations of two modalities, improving the quality of aligned representation, which is crucial for downstream tasks such as classification and regression.

\textbf{Annotation-Protein Matching (APM) Loss}. Inspired by ~\cite{li2022blip}, we introduce a multi-modal encoder $f_{\text{ME}}$, which integrates the representation of annotations and protein as $\bm{z}_{S, \mathcal{A}}^{M} = f_{\text{ME}}(S, G(\mathcal{A}))$, to identify whether the given pairs of protein description $G(\mathcal{A})$ and protein $S$ are matched or not. Specially, the multimodal encoder $f_{\text{ME}}$ is transformer-based and shares the parameters of self-attention and feed-forward layers with $f_{\text{PE}}$ and has additional cross-attention layers between self-attention layers and feed-forward layers to integrate the text information. The cross-attention layers share the same cross-attention parameters as the decoder.

Annotation-Protein Matching (APM) Loss aims to facilitate the learning of multimodal representations. %
Additionally, we use the hard negative strategy, where we first select the most similar negative pairs and then use these most challenging negative pairs to optimize the model through
in the training of the model with the APM loss, enabling the encoder to learn informative representations. To construct the APM loss, we extract the multi-modal representation as $\bm{z}_{S, \mathcal{A}}^{M} = f_{\text{ME}}(S, G(\mathcal{A}))$ and use a classifier to classify its probability of being positive or negative, which is denoted by $\Vec{p}^\textrm{APM}(\mathcal{A}, S)$. And we compute the cross-entropy between the ground-truth label $\Vec{y}^\textrm{APM}(A,S)$ indicating the protein-description pair being positive or negative to obtain the APM loss as
\vspace{-1ex}
\begin{equation}
\vspace{-1ex}
\label{eqn:APM}
\mathcal{L}_\mathrm{APM} = \mathbb{E}_{(A,S)} \mathrm{H} (\Vec{y}^\textrm{APM}(A,S), \Vec{p}^\textrm{APM}(A,S)).
\end{equation}

APM loss trains the cross-attention layers for integrating two modalities. These cross-attention will be shared with protein decoder, which helps protein decoder interpret textual information. In Sec.~\ref{sec.abstudy}, we demonstrate the importance of APM loss.

\subsection{Conditional Protein Decoding}
\label{Conditional Protein Decoding}

\textbf{Protein Decoder. }
Our ultimate goal is generating functional proteins conditioned on a set of annotations $\mathcal{A} = \{T_k\}_{k=1}^{K}$. 
We adopt an auto-regressive protein decoder $f_{\text{D}}$ that can receive the condition from the language model. Specifically, the decoder will first process the protein sequence through causal attention layers to enable auto-regressive generation, then integrate the information from annotations via cross-attention layers followed by feed-forward layers. Note that the causal attention layers are initialized using the same weights as the self-attention layers in the protein encoder and the feed-forward layers share the same parameters as the protein encoder. Here, the parameter-sharing mechanism enables higher training efficiency~\cite{li2023blip}. The cross-attention layers are randomly initialized and trained from scratch. 

\textbf{Protein Modeling (PM) Loss}. To guide the learning of the model, we estimate the PM loss as 
\vspace{-1.5ex}
\begin{equation}
\vspace{-1.5ex}
 \mathcal{L}_{PM} = - \sum_{i=1}^{l} \log {p(S_{i} | S_{<i}; \mathcal{A})},
\end{equation}
where $l$ is the length of the protein $S$ and $p(S_{i} | S_{<i}; \mathcal{A})$ is the predicted probability of the $i$-th amino acid given all previous amino acids and the annotation set.

\subsection{Training Objectives}
\label{Training Objectives}
In the training of {\method}, we optimize four objective functions. These functions are designed to align the representations between annotations and protein sequences, integrate the two modalities, and reconstruct the protein sequence. In contrast to ProteinDT~\cite{liu2023textguided}, which splits the training process into three separate stages, {\method} jointly optimizes these four objective functions in an end-to-end manner.
The overall pretraining objective of {\method} is :
\vspace{-1ex}
\begin{equation}
\vspace{-1ex}
\mathcal{L} = \mathcal{L}_\mathrm{ADC}  + \mathcal{L}_\mathrm{APC} + \mathcal{L}_\mathrm{APM} + \mathcal{L}_\mathrm{PM}.
\end{equation}

\subsection{Annotation-guided Protein Design}
After obtaining the model $F$ of {\method}, we can use $F$ to design the proteins with given textual annotations. In the following, we given the definition of annotation-guided protein design.

\begin{definition}[Annotation-guided Protein Design]
Given an annotation set $\mathcal{A} = \{T_k\}_{k=1}^{K}$ with $K$ annotations, a generative model $F$ can leverage the information from $\{T_k\}_{k=1}^{K}$ to generate the protein $S$ which satisfies the condition described in $\{T_k\}_{k=1}^{K}$. Quantitatively, this task aims to maximize the following objective function:
\vspace{-1.5ex}
\begin{align}
\vspace{-1.5ex}
    \max \sum_{T \in \mathcal{A}}\mathrm{M}_{T}(S), \text{ s.t. } S=F(\mathcal{A}),
\end{align}
where $\mathrm{M}_{T}(\cdot)$ is metric function for the annotation $T$. 
\end{definition}
In this paper, we mainly focus on two type of metrics, \emph{functional-domain metric} and \emph{global-property metric}. 
\begin{itemize}[leftmargin=*]
\item \textbf{Functional-domain metric}: for the annotation describing a certain functional domain, $\mathrm{M}_{T}()$ will invoke a profile hidden Markov models from Pfam~\cite{mistry2021pfam} to search for the optimal match within protein $S$ regarding to the domain described by annotation $T$ and assign an e-value $s$ to this match. Given an e-value threshold $e$, we have:
\vspace{-1ex}
\begin{align}
\vspace{-2ex}
    \mathrm{M}_{T,e}(S) =  \mathds{1}_{S}(s < e),
    \label{equ:mts}
\end{align}
where $\mathds{1}_{S}(\text{cond})$ outputs $1$ when $\text{cond}=\text{True}$, otherwise, outputs $0$. 
\item \textbf{Global-property metric}: for the annotations describing protein's global properties, $\mathrm{M}_{T}()$ employ a pre-defined oracle to determine if the given $S$ has the property described by the annotation $T$ and output a  score $s$ in the range of $[0,1]$. Here, 0 denotes the absence of the property, while 1 signifies its presence. 
\end{itemize}

\section{Experiment}
\label{Experiment}

In this section, we extensively evaluate {\method} from two aspects: (1). quality of aligned protein representation for the predictive tasks; (2) evaluation on the unconditional sequence generation and protein design with textual annotations.

\subsection{Construction of {\dataset} Dataset} 
To enable multi-level alignment, we build the {\dataset} dataset with annotation-sequence pair set for protein design task. Specifically, we select the proteins with ``Domain'' entry in UniProtKB~\cite{10.1093/nar/gkac1052} to build {\dataset} dataset.

However, since the biases within our training dataset will significantly impact the performance of {\method}, we limit our selection to protein-annotation pairs from a refined subset of UniProtKB, namely Swiss-Prot. Each entry in Swiss-Prot has been manually reviewed and supplemented with detailed information with protein function, structure, and interactions.
Ultimately, the {\dataset} consists of 129,727 proteins. 
Moreover, in our alignment, we incorporate temperature \cite{wu2018unsupervised} to control the similarity scores of distribution between positive and negative pairs. An appropriate temperature can soften these scores, making the model less sensitive to noisy samples. This adaptability allows the learning process to focus more on meaningful correlations and reduce the impact of irrelevant variations, ultimately enhancing the PAAG’s robustness.

The domain annotations are extracted by these ``Domain'' entries including the domain description and start \& end index of this domain.  Additionally, we select four properties as the property annotations, that is, ``protein\_name'', ``organism\_name'', ``length'' and ``SIMLARITY''. The textual description is assembled by the template function $G$, which can accept any subset of annotation as input. More details, including the data examples are deferred in Appendix~\ref{appendix.dataset}.

\subsection{Quality of Aligned Representation}
\label{Quality of Aligned Representation}
We first conduct multiple experiments on predictive tasks, 
to evaluate the quality of protein representation produced by {\method}.

\textbf{Settings:} To make a fair comparison with ProtST, we use the same proteins in dataset ProtDescribe~\cite{xu2023protst} to first pretrain {\method}, followed by full-model fine-tuning on various downstream tasks. For the full-model fine-tuning, we add a task head for each task and fine-tune the model for 100 epochs. We use the validation set to select the model and report the results on random seed 0,  adhering to the same settings as in ProtST~\cite{xu2023protst}. More details are deferred in Appendix~\ref{appendix.training_configurations}. 

\textbf{Benchmark Tasks:} 
We adopt 7 downstream tasks within two task types as the benchmark task.
\begin{itemize}[leftmargin=*]
    \item \textbf{Protein Localization Prediction} aims to forecast the subcellular locations of proteins. Derived from DeepLoc~\cite{almagro2017deeploc}, we focus on two similar tasks. The subcellular localization prediction (\textit{Abbr. as}, Sub) encompasses 10 location categories and binary localization prediction (\textit{Abbr. as}, Bin) that includes 2 location categories, soluble and membrane-bound. The splits of data follow the original split in DeepLoc. \eat{The detailed statistics of these two datasets are in Appendix ~\ref{appendix: dataset stats}.}
    \item \textbf{Fitness Landscape Prediction} is primarily focused on the prediction of the effects of residue mutations on the fitness of proteins. We evaluate our models on $\beta$-lactamase (\emph{Abbr.}, $\beta$-lac) landscape from PEER~\cite{xu2022peer}, the AAV and Thermostability (\emph{Abbr.}, Thermo) landscapes from FLIP~\cite{dallago2021flip}, and the Fluorescence (\emph{Abbr.}, Flu) and Stability (\emph{Abbr.}, Sta) landscapes from TAPE~\cite{rao2019evaluating}. The splits of data follow the splitting setting of ProtST~\cite{xu2023protst}
\end{itemize}

\textbf{Baselines:} We adopt two types of baseline. The first type is the models trained from scratch, including CNN~\cite{shanehsazzadeh2020transfer}, ResNet~\cite{rao2019evaluating}, LSTM~\cite{rao2019evaluating}, Transformer~\cite{rao2019evaluating}. The second type is the pre-trained models with full model tuning, including OntoProtein~\cite{zhang2022ontoprotein}, ProtBert~\cite{elnaggar2021protrans} and {ESM2~\cite{lin2023evolutionary}}. For the ProtST and {\method}, we train two variants with different initialization weights from ProtBert and ESM2 to verify the enhancement of text knowledge on protein representations from different protein encoders. We utilize distinct subscripts to denote the initialized parameters, such as $\text{\method}_{\text{ProtBert}}$ and $\text{\method}_{\text{ESM2}}$.

\begin{table}[t]
  \centering
  \caption{Results on protein localization prediction and protein landscape prediction benchmarks. Bold in green and underlined numbers indicate the best and the second best result, respectively. }
  \vspace{-2ex}
  \resizebox{\columnwidth}{!}{%
\begin{tabular}{l|rr|rrrrr}
\toprule
\multirow{2}[4]{*}{\textbf{Model}} & \multicolumn{2}{c|}{\textbf{Loc. pred.} (Acc\%)} & \multicolumn{5}{c}{\textbf{Fitness pred.} (Spearman's $\rho$)} \\
\cline{2-8}           & \multicolumn{1}{c}{\textbf{Bin} $\uparrow$} & \multicolumn{1}{c|}{\textbf{Sub} $\uparrow$} & \multicolumn{1}{c}{\textbf{$\rho$-lac} $\uparrow$} & \multicolumn{1}{c}{\textbf{AAV} $\uparrow$} & \multicolumn{1}{c}{\textbf{Thermo} $\uparrow$} & \multicolumn{1}{c}{\textbf{Flu} $\uparrow$} & \multicolumn{1}{c}{\textbf{Sta} $\uparrow$} \\
\midrule
\multicolumn{8}{c}{\textbf{Models trained from scratch}} \\
\midrule
\textbf{CNN}        & 82.67      & 58.73      & 0.781      & 0.746      & 0.494      &  \cellcolor[rgb]{ .757,  .941,  .784}\textbf{0.682} & 0.637 \\
\textbf{ResNet}     & 78.99      & 52.30      & 0.152      & 0.739      & 0.528      & 0.636      & 0.126 \\
\textbf{LSTM}       & 88.11      & 62.98      & 0.139      & 0.125      & 0.564      & 0.494      & 0.533 \\
\textbf{Transformer} & 75.74      & 56.02      & 0.261      & 0.681      & 0.545      & 0.643      & 0.649 \\
\midrule
\multicolumn{8}{c}{\textbf{Models with full model tuning}}\\
\midrule
\textbf{OntoProtein} & 92.47      & 77.59      & 0.757      & 0.791      & 0.662      & 0.630      & 0.731 \\
\textbf{ProtBert}   & 91.32      & 76.53      & 0.731      & 0.794      & 0.660      & 0.679      & 0.771 \\
\textbf{ESM2}      & 91.72      & 78.67      & 0.867      & 0.817      & 0.672      & 0.677      & 0.718 \\
$\textbf{ProtST}_\text{ProtBert}$ & 91.78      & 78.71      & 0.863      & 0.804      & 0.673      & 0.679      & \underline{0.745} \\
$\textbf{ProtST}_\text{ESM2}$ & \underline{92.52}      & \underline{80.22}      & \underline{0.879}      & \underline{0.825}      & \underline{0.682} &  \cellcolor[rgb]{ .757,  .941,  .784}\textbf{0.682} & 0.738 \\
\midrule
$\textbf{PAAG}_\text{ProtBert}$ &  \cellcolor[rgb]{ .757,  .941,  .784}\textbf{92.63} & 78.96      & 0.820       & 0.825      & 0.668      & \underline{0.680}      & \cellcolor[rgb]{ .757,  .941,  .784}\textbf{0.788 } \\
$\textbf{PAAG}_\text{ESM2}$  & 92.46      &  \cellcolor[rgb]{ .757,  .941,  .784}\textbf{81.30} &  \cellcolor[rgb]{ .757,  .941,  .784}\textbf{0.888} &  \cellcolor[rgb]{ .757,  .941,  .784}\textbf{0.839} & \cellcolor[rgb]{ .757,  .941,  .784}\textbf{0.684}      &  \cellcolor[rgb]{ .757,  .941,  .784}\textbf{0.682} & 0.737 \\
\bottomrule
\end{tabular}%
}
\vspace{-3ex}
  \label{tab:predictive_res}%
\end{table}%

\textbf{Results:} 
Table~\ref{tab:predictive_res} reports the results of all models on seven baselines. As illustrated in Table~\ref{tab:predictive_res}, {\method} achieves superior performance in comparison to the baselines on all 7 tasks. These results highlight the robust generalization capabilities of {\method} for downstream tasks. Specifically, {\method} outperforms the vanilla pretrained models, ProtBert and ESM2, in all cases, indicating that {\method} can further enhance the quality of protein representation by incorporating the knowledge from textual annotations. \eat{Furthermore, to assess the impact of multimodal training on enhancing protein representations, we visualize the relative improvements of ProtST and {\method} in comparison to the vanilla pre-trained models in Figure~\ref{fig:prediction}. As observed in Figure~\ref{fig:prediction},} 
According to the results in Table~\ref{tab:predictive_res}, multi-modal training has a positive influence on the performance of downstream tasks in both ProtST and {\method}. Additionally, the performance improvement of {\method} surpasses that of ProtST in 10 out of 14 cases, further validating the effectiveness of {\method} in augmenting the quality of protein representation.

\subsection{Unconditional Protein Generation}
To verify the learning effect of the decoder, we compare the ability of different models in unconditional generation task. A good decoder is able to generate protein sequences that conform to the distribution of the training set while simultaneously exhibiting adequate novelty. 

\textbf{Setting:} In unconditional generation task, we only specify the length of generated proteins and conditions. We sample the same length from natural proteins, to ensue a fair comparisons across different models.

\textbf{Baselines:}
We compare {\method} with 3 representative protein design models, \textit{i.e.}, ProGen ~\cite{madani2020progen}, Chroma ~\cite{ingraham2022chroma} and ProteinDT ~\cite{liu2023proteindt}. Furthermore, we introduce two naive baselines: Random$_\text{Uniform}$ and Random$_\text{Empirical}$. Random$_\text{Uniform}$ generates protein sequence by randomly selecting amino acids based on a uniform distribution, while Random$_\text{Empirical}$ adheres to the empirical amino acid distribution in the training dataset. Additionally, we report the results of the sequence set sampled from natural proteins, denoted as Natural, to serve as a reference. The details of baselines are in Appendix~\ref{appendix.baselines}.

\textbf{Evaluation metrics:}
To evaluate the quality of generated protein sequences, we employ three metrics: Distinct-n, Diversity and Novelty. Suppose $\mathcal{S}$ is the protein sequence set. 
\begin{itemize}[leftmargin=*]
    \item \textbf{Distinct-n}~\cite{li-etal-2016-diversity} is a classical metric in natural language processing that measures textual diversity of generated text by counting distinct n-grams. We use normalized Distinct-n to assess the fraction of repetitive sequence motifs in sequences from $\mathcal{S}$, which exhibits the biological importance~\cite{andrade2000homology}. A higher Distinct-n suggests fewer repetitive amino acid segments. We set $n=2$ here. 
    \item \textbf{Diversity} measures the dissimilarity of sequences in  $\mathcal{S}$. We employ Mmseq2~\cite{steinegger2017mmseqs2} to compute the dissimilarity between each pair of sequences, and utilize the mean of these dissimilarities as Diversity. A higher Diversity signifies a greater diversity in $\mathcal{S}$. 
    \item \textbf{Novelty} measures the novelty of sequences in  $\mathcal{S}$ compared to a reference set. We take UniprotKB ~\cite{10.1093/nar/gkac1052} as the reference set. For each sequences in $\mathcal{S}$, we employ Mmseq2~\cite{steinegger2017mmseqs2} to return the dissimilarity score of the most similar sequence in UniprotKB. Novelty is defined as the mean of dissimilarity score for all sequences in $\mathcal{S}$. A higher Novelty indicates the generated sequences exhibit substantial novelty comparing with the reference set. 
\end{itemize}

\eat{
We also introduce additional two metrics, novelty and diversity. Intuitively, novelty assesses whether there are highly similar proteins exist in the natural protein dataset, and diversity evaluates whether exist similar proteins within the generated protein sequences. Therefore, we utilize an efficient search tool, Mmseq2\cite{steinegger2017mmseqs2}, which can identify the similar sequences within a given protein database. We search our generated proteins in the UniprotKB ~\cite{10.1093/nar/gkac1052} and define novelty as:

\begin{equation}
 Novelty = \frac{\sum_{i=1}^M \mathds{1}_{P_i}(e\text{-}value_{P_i} < T)A_{UniprotKB}}
 {\sum_{i=1}^M \mathds{1}_{P_i}(e\text{-}value_{P_i} < T)}
\end{equation}
where $A_{UniprotKB}$ is the max alignment score of each generated proteins searched in UniprotKB.

\begin{equation}
 Diversity = \frac{\sum_{i=1}^M \mathds{1}_{P_i}(e\text{-}value_{P_i} < T)A_{Others}}
 {\sum_{i=1}^M \mathds{1}_{P_i}(e\text{-}value_{P_i} < T)}
\end{equation}
where A is the max alignment score of each generated proteins searched in other generated proteins.
}

\begin{table}[t]
  \centering
  \caption{Results of the unconditional generation task.  The values in parentheses $\Delta \%$ represent the absolute relative difference from the values of Natural.}
  \vspace{-2ex}
  \resizebox{0.8\linewidth}{!}{
    \begin{tabular}{l|rrr}
    \toprule
    \textbf{Model} & \multicolumn{1}{l}{\textbf{Distinct-2} ($\Delta \%$)} & \multicolumn{1}{l}{\textbf{Diversity}($\Delta \%$)} & \multicolumn{1}{l}{\textbf{Novelty} $\uparrow$} \\
    \midrule
    \textbf{Natural} & 0.4309(0) &   0.829(0)   & - \\
    \midrule
    \textbf{Random}$_\text{Uniform}$ & 0.5006(16.18\%) & 0.847(2.17\%) & 0.713 \\
    \textbf{Random}$_\text{Empirical}$ & 0.4442(3.09\%) & 0.834(\textbf{0.60\%}) & 0.721 \\
    \midrule
    \textbf{ProGEN} & 0.3003(30.31\%) & 0.845(1.93\%) & 0.374 \\
    \textbf{Chroma} & 0.3211(25.48\%) & 0.855(3.14\%) & 0.638 \\
    \textbf{ProteinDT} & 0.4909(13.92\%) & 0.814(1.81\%) & 0.578 \\
    \midrule
    \textbf{PAAG}  & 0.4314(\textbf{0.12\%}) & 0.815(1.69\%) & \textbf{0.766} \\
    \bottomrule
    \end{tabular}%
}
\vspace{-5ex}
  \label{tab:unconditional}%
\end{table}%

\begin{figure*}[th]
    \centering
    \includegraphics[width=0.24\textwidth]{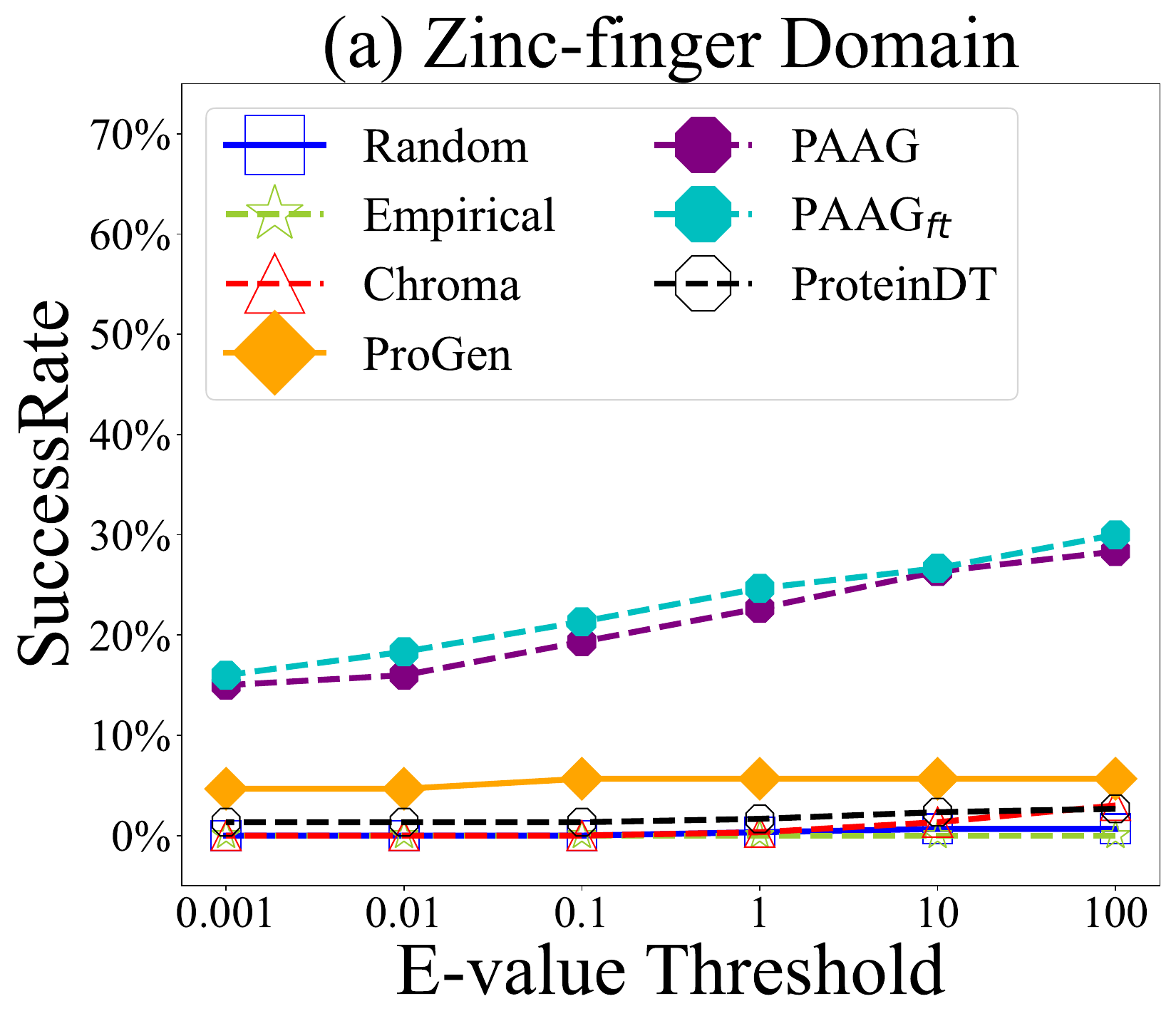}
    \includegraphics[width=0.24\textwidth]{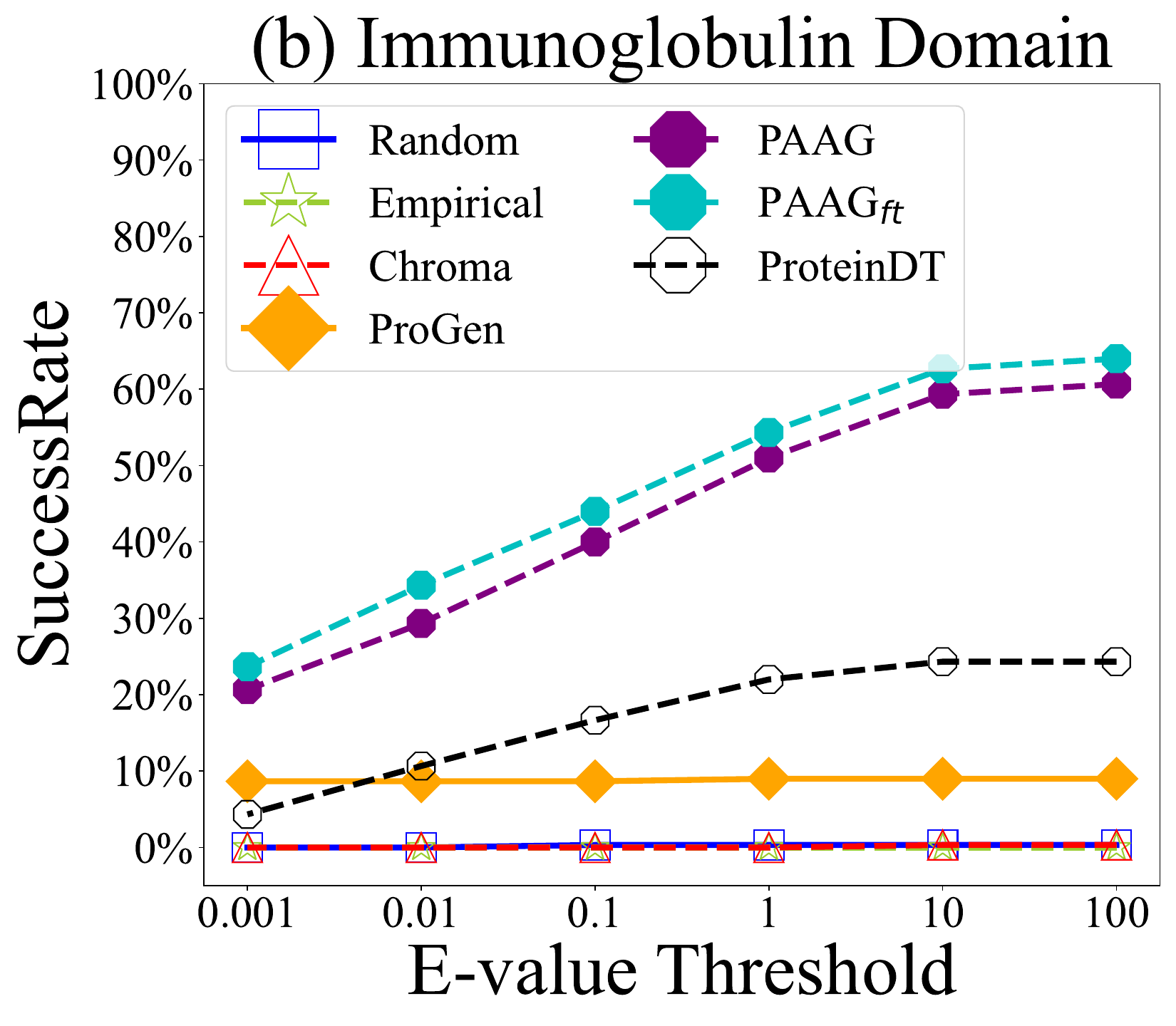}
    \raisebox{-2.3ex}{\includegraphics[width=0.24\textwidth]{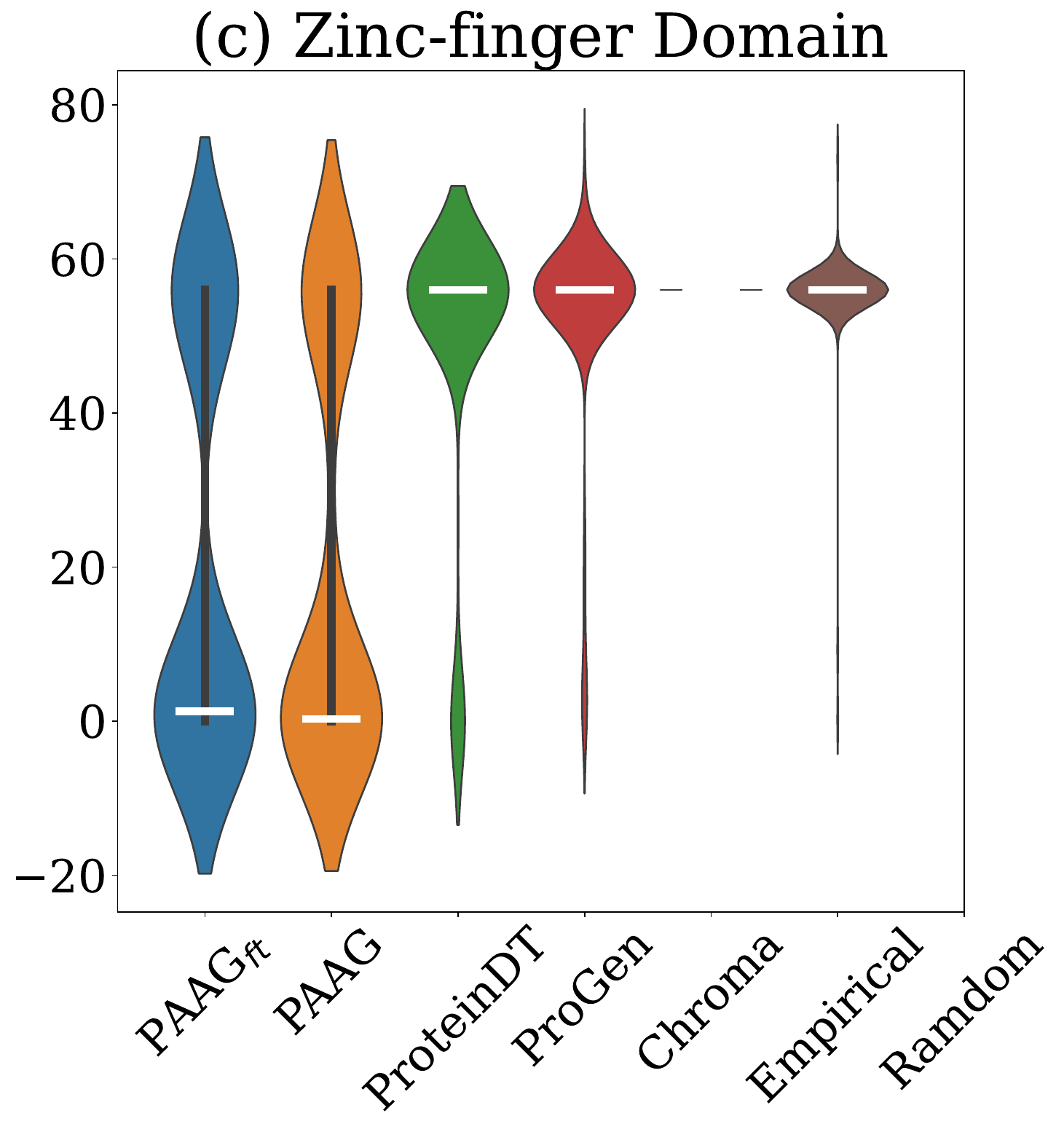}}
     \raisebox{-2.3ex}{\includegraphics[width=0.24\textwidth]{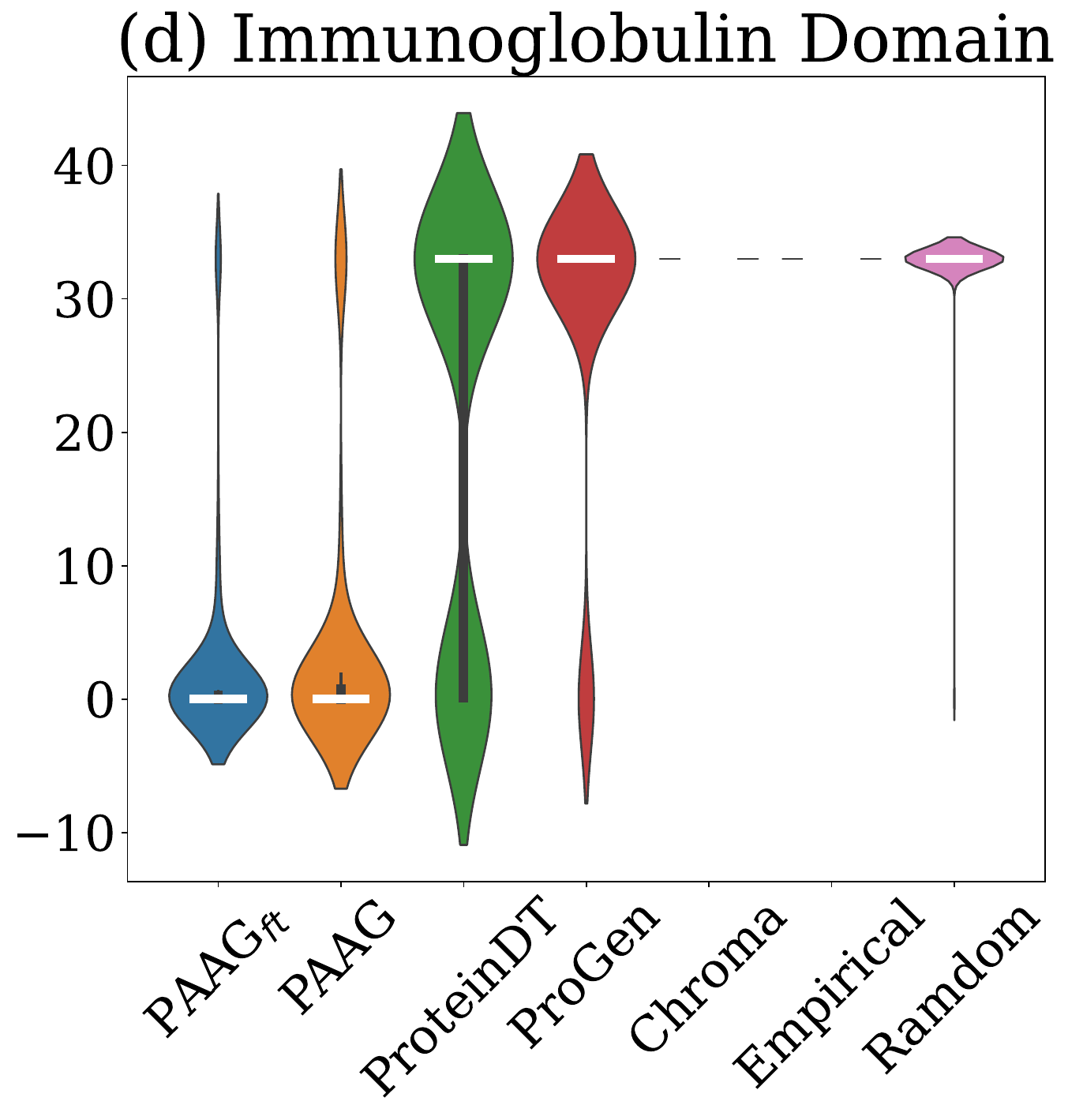}}
    \vspace{-4ex}
    \caption{Figure (a) and (b) show the $\text{SR}_{e}$ on zinc-finger domain and immunoglobulin domain over all models. Figure (c) and (d) show their distributions of e-value. White bar indicates the mean e-value of each set. {\method} consistently exhibits better performance on all metrics compared with other models. Fine-tuning also introduces additional improvement for {\method}.
    }
    \vspace{-3ex}
    \label{fig:conditional}
\end{figure*}

\begin{figure*}[thb]
    \centering
    \includegraphics[width=0.22\textwidth]{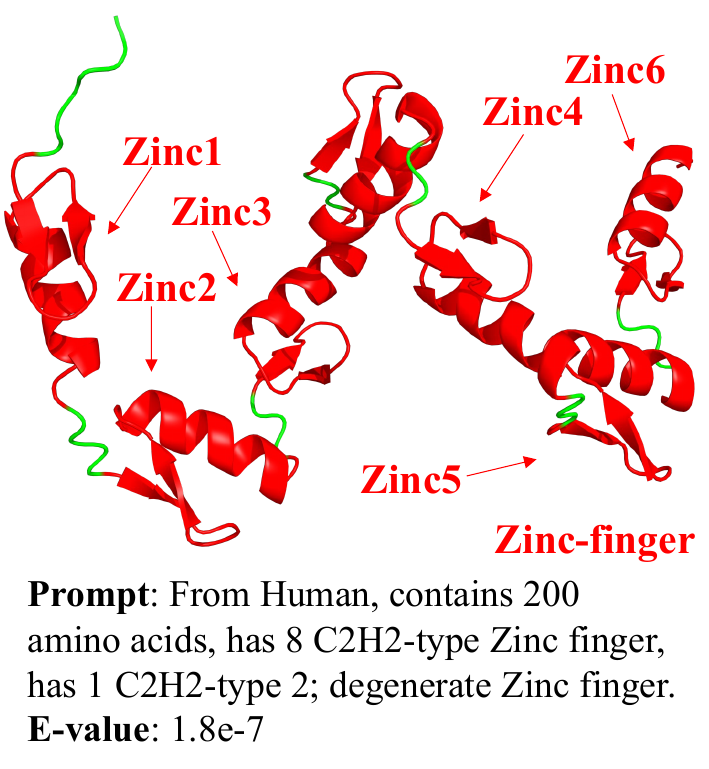}
    \includegraphics[width=0.22\textwidth]{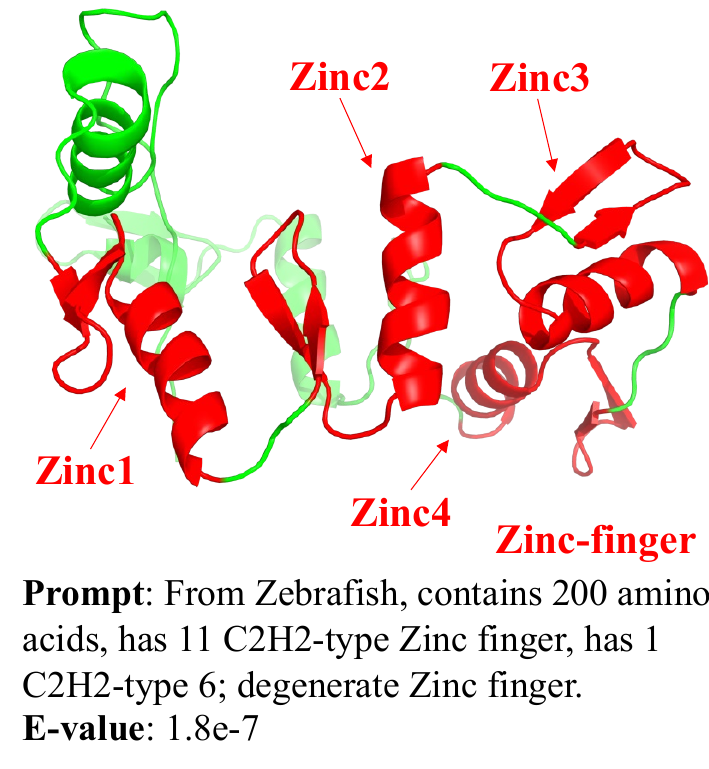}
    \includegraphics[width=0.22\textwidth]{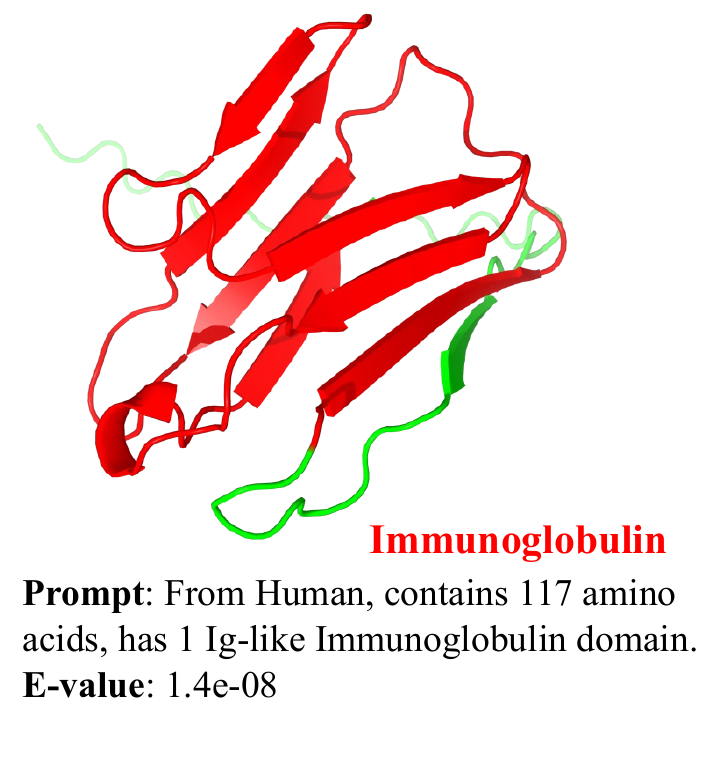}
    \includegraphics[width=0.22\textwidth]{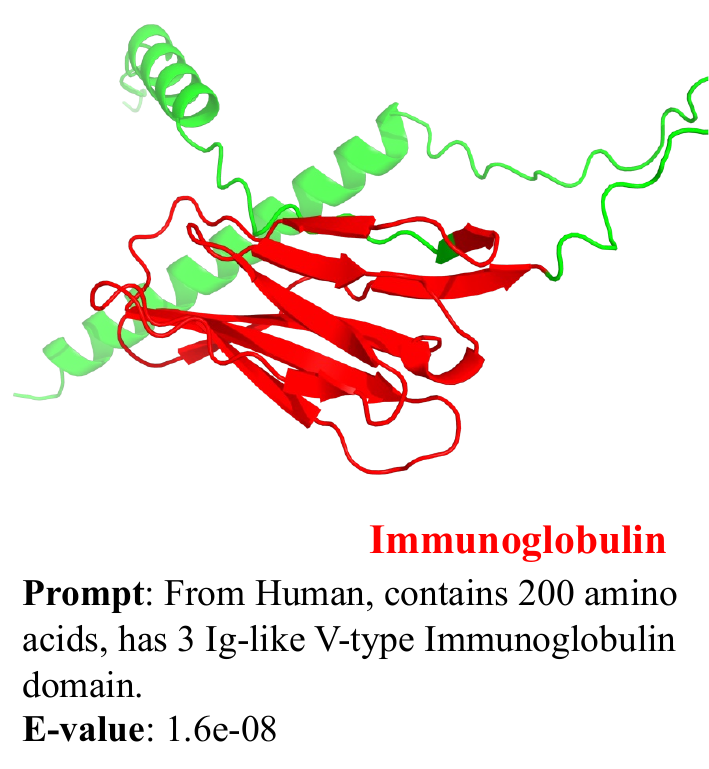}
    \vspace{-3ex}
    \caption{Visualization of the generated results on zinc-finger and immunoglobulin domain. The corresponding prompt and generation qualify (e-value) is listed below. 
    }
    \vspace{-4ex}
    \label{fig:visualization}
\end{figure*}

\textbf{Results:}
Table ~\ref{tab:unconditional} shows the results of unconditional generation under three metrics. From Table ~\ref{tab:unconditional}, we can observe that:
\begin{itemize}[leftmargin=*]
    \item {\method} achieves the highest novelty, indicating {\method} captures the intrinsic relationship between amino acids, rather than merely memorizing protein sequences in the training set.
    \item  {\method} exhibits the closest Distince-n score compared with natural proteins, which proves that proteins generated by {\method} possess similar amino acid distribution as natural proteins.
    \item Since Random$_\text{Empirical}$ generates protein sequence following the empirical amino acid distribution in the natural proteins, it is reasonable to obtain the closest diversity to natural proteins. Compared with other learning-based model, {\method} still maintains the closest diversity to natural proteins proving {\method} has the similar distribution at protein level. 
    \item ProGen and Chroma have considerably lower Distince-n score in comparison to {\method} and natural proteins, implying an abundance of repetitions within generated protein sequence. While ProteinDT has higher Distince-n score, it also fails to capture the intrinsic amino acid distrbution in natrual proteins.
\end{itemize}
In a summary, {\method} is capable of generating high-quality protein sequences, aided by the the multi-level alignment process.

\subsection{Protein Design with Domain Annotations}
\label{Protein Design with Domain Annotations}
In this section, we evaluate the performance of {\method} in generating proteins under the given domain annotations. 

\textbf{Settings:}
We utilize two biologically significant domains, zinc-finger domain~\cite{klug1987zinc} and immunoglobulin domain~\cite{brummendorf1995cell}, as the target domain annotation to generate the proteins respectively. For each case, we generate $N=300$ protein sequences given the length of the proteins and the annotation set containing the domain annotations of ``zinc-finger'' or ``immunoglobulin domain''. For all models, the protein sequences are generated with the same length sampled from natural proteins that have corresponding domain in UniprotKB. To further evaluate the generalization ability of PAAG, we further test our model on generating proteins with EGF-like domain. The details can be found in Appendix~\ref{appendix.egf}.

Based on \eqref{equ:mts}, we further define a metric: success rate $\text{SR}_{e}=\frac{M_{T, e}(S)}{N}$ to measure the generation quality with the proportion of proteins that successfully identify the specific domain with the quality threshold $e$. 

\textbf{Model training:} We train {\method} on {\dataset} for total 100 epochs, employing a learning rate of 3e-5 and incorporating a warm-up phase with a batch size of 32.
Additionally, we extract a subset of {\dataset} which only contains proteins with ``zinc-finger'' and ``immunoglobulin domain''. This subset is subsequently utilized to fine-tune the model over 5 epochs, adopting a learning rate of 1e-5 and maintaining the batch size at 32. We denote this fine-tuned version as $\text{\method}_{\text{ft}}$, and the PAAG without fine-tuning is denoted as PAAG. The generalization ability of $\text{\method}_{\text{ft}}$ is also evaluated in ~\ref{appendix.evaluationmetric}. More details are deferred to {Appendix~\ref{appendix.fine-tuned unconditional}}.

\textbf{Baselines:}
We adopt ProtenDT~\cite{liu2023proteindt}, Chroma~\cite{ingraham2022chroma} and ProGen~\cite{madani2020progen} as the baselines due to their public availability of model weights and their capacity to accept text and keywords as conditional inputs for protein generation.
ProGen utilizes keywords as its condition tags. Here, we take the keywords correspond to zinc-finger (``KW-0863'') and immunoglobulin domain (``KW-0393'') in Uniport to generate functional proteins. Chroma adopts ProCap to understand the textual conditions to guide its diffusion process. We give the same text prompts to Chroma to enable its controllable generation. We also include two trivial baselines Random$_\text{Uniform}$ and Random$_\text{Empirical}$ as the blank references. 

\textbf{Results:} 
In Figure~\ref{fig:conditional} (a) and (b), we plot the curves depicting the variation in success rate $\text{SR}_{e}$ for different models in two protein design tasks as the quality threshold $e$ varies from 0.001 to 100. Additionally, in Figure~\ref{fig:conditional} (c) and (d), we employ violin plots to illustrate the distribution of e-value $e$ for generated sequences matched in Pfam. Due to Pfam's limitations, e-values exceeding 100 are adjusted to the maximum e-value score among all method results for the current task. We also show the distribution of e-values calculated on natural proteins with domains as a reference. 
From Figure~\ref{fig:conditional}, we can observe that:
\begin{itemize}[leftmargin=*]
    \item {\method} achieves significantly higher success rates $\text{SR}_1$ across all tasks by a large margin, e.g. , $51\%$ versus $22\%$ in immunoglobulin domain task with quality threshold $e=1$. Furthermore, the fine-tuned $\method_{ft}$ can further improve the success rates (54.3\% on $\text{SR}_1$) consistently, indicating the finetuning process can help to further improve the quality.
    \item ProGen and ProteinDT outperform other baseline methods. This may be due to they memorizing proteins with zinc and lg domain in the training data and output them when using keywords or textual prompts. 
    \item Chroma's performance is not good, resembles the unconditioned results.
     This may be because Chroma's training text primarily focuses on structural descriptions, making it less sensitive to the domain annotations.    
\end{itemize}

\vspace{-1ex}
\textbf{Visualizations: } We provide the visualization of generated proteins, folded by Omegafold~\cite{wu2022high}, in Figure ~\ref{fig:visualization}. 
The figure ~\ref{fig:visualization} highlights generated domains in red and provides textual descriptions and e-value $e$ for each sequence.
We observe the generated sequences accurately produce target domains as specified in the annotation set. Interestingly, in scenarios such as two zinc finger cases, when the prompt specifies the presence of multiple zinc finger domains, {\method} generates multiple functional domains in response. However, {\method} fails to capture the precise numbers of these domains, which can be a direction for future improvement. 
\eat{Further analysis of the relationship between generated and prompt domains is provided in the Appendix ~\ref{appendix.numberofdomains}.
We also assess the success rate on prompts contain only single domain in the Appendix~\ref{appendix.singledomainsuccess}}

\begin{figure}[t]
\begin{center}
\centerline{\includegraphics[width=0.6\linewidth]{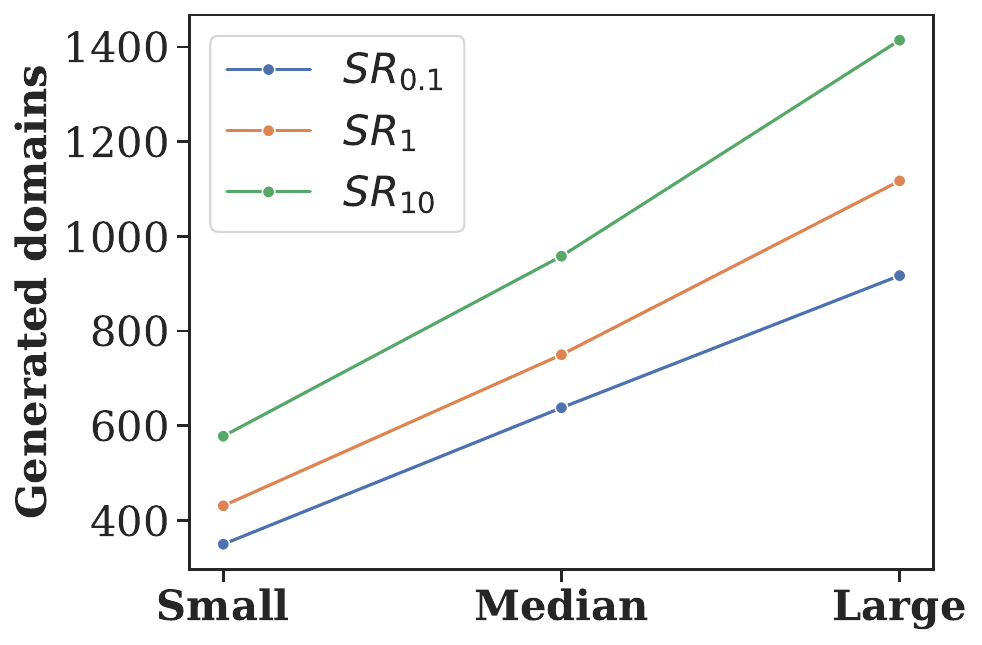}}
\vspace{-3ex}
\caption{The relation of number specified in prompt with generated domains by {\method}. \eat{Small, Medium and Large indicate there are 1-3, 4-6 or 7-9 zinc-finger domains specified in prompts.}}
\label{fig:prompt_number}
\end{center}
\vspace{-6ex}
\end{figure}

\textbf{Prompt}: 
We further investigate the relationship between the number of domains in prompts and the proteins generated by {\method}. By varying only the domain count in prompts, we generate 900 proteins with `Small (1-3)', `Median (4-6)', and `Large (7-9)' number of domains. Figure~\ref{fig:prompt_number} shows that increasing domain numbers in prompts leads to a corresponding rise in generated domains across different e-values, indicating that {\method} can discern and generate the specified number of domains from text prompts through multi-level alignment. 
Given these observations, another interesting question is whether multiple domains in prompt will increase success rates. To this end, we re-evaluate PAAG using prompts specified only one domain (PAAG$_{\text{single domain}}$). The results are in Table \ref{tab:singledomain} in Appendix. We can find while multiple domains improve success rates, PAAG$_{\text{single domain}}$ also outperforms the baselines, confirming the importance of multi-level alignment.

{

}

\eat{
\textbf{Case study: Multiple Functional Domain Protein Generation}

In the natural world, there is no protein that simultaneously possesses both zinc finger and immunoglobulin domain. Consequently, we can demonstrate the generalized ability of {\method} if {\method} can generate a protein that contains both zinc-finger domain and immunoglobulin. Furthermore, we can also show the flexibility of forming the textual condition by simply editing the text.
}

\subsection{Protein Design with Property Annotations}
\label{sec.prop_anno_gen}
In this section, we explore the potential of {\method} to generate proteins with certain properties guided by property annotations. 

\textbf{Settings: }  We employ the subcellular location of proteins as an example property. An additional dataset, termed ProtLocation, is generated by extracting subcellular location labels, encompassing Bin (2-class) task as delineated in Section \ref{Quality of Aligned Representation}, from Deeploc~\cite{almagro2017deeploc}. These labels are then incorporated into annotation-sequence pairs derived from Uniprot. ProtLocation includes $10100$ proteins for training, while a separate set of $2434$ test proteins is reserved for constructing the annotation set for generation. More details can be found in Appendix~\ref{appendix:protannotation config}.

\textbf{Evaluation protocol:} For the generated sequences, we employ the official server provided by Deeploc~\cite{almagro2017deeploc} as the pre-defined oracle to construct the Global-property metric $\mathrm{M}_{T}()$ for predicting binary location label. A generation is deemed successful if the predicted label aligns with the input annotation label.

\textbf{Results: }  As shown in Table \ref{tab:bin_anno_result}, {\method} achieves overall $74.78\%$ success rate, indicting {\method} captures the difference between soluble and membrane-bound, and can generate properties given corresponding property annotations. 
\eat{More results are in Appendix \ref{appendix:predicted probabilities}. }
We next will move to more challenging setting, generating proteins with both domain and property annotations.

\begin{table}[h]
  \centering
  \vspace{-2ex}
  \caption{The result of protein design with   ``membrane-bound'' and ``soluble'' annotations.}
  \vspace{-2ex}
  \resizebox{0.75\linewidth}{!}{
    \begin{tabular}{lrrr}
    \toprule
    \textbf{Annotation}   & \textbf{Total}      & \textbf{Matched}    & \textbf{Success Rate} \\
    \midrule
    \textbf{membrane-bound} &    660      &    351     &  53.18\% \\
    \textbf{soluble}    &      894    &     811    &    90.72\%\\
    \textbf{overall}    &  1554      & 1162 & 74.78\% \\
     \bottomrule
    \end{tabular}%
    }
    \vspace{-3ex}
  \label{tab:bin_anno_result}%
\end{table}%
\begin{figure*}[t]
    \centering
    \includegraphics[width=0.23\textwidth]{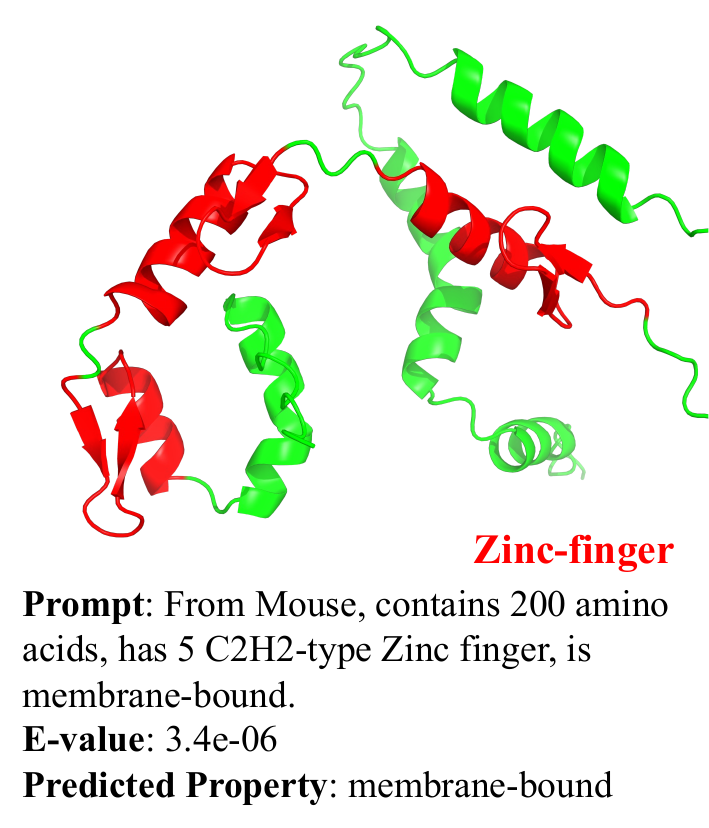}
    \includegraphics[width=0.23\textwidth]{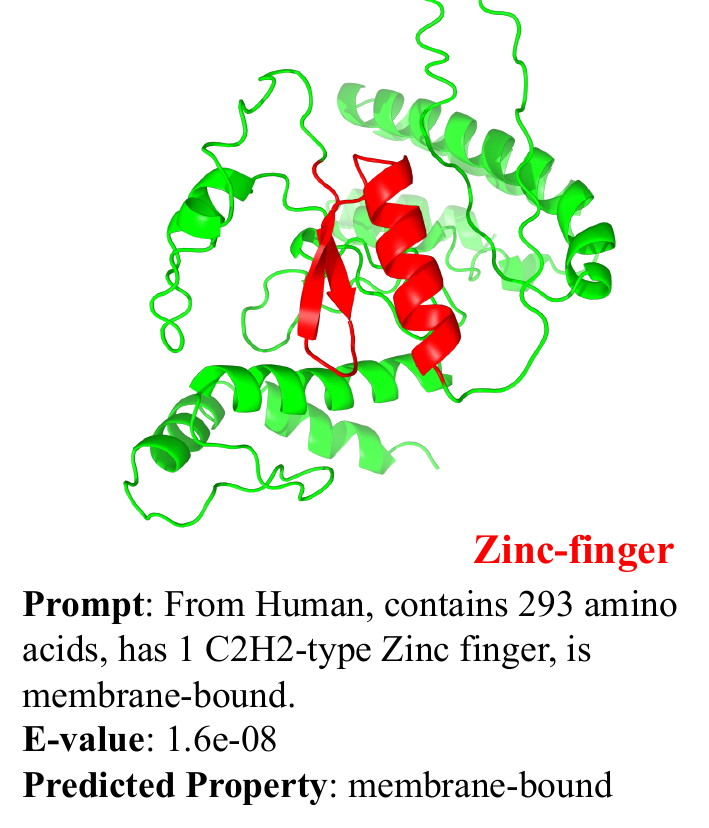}
    \includegraphics[width=0.23\textwidth]
    {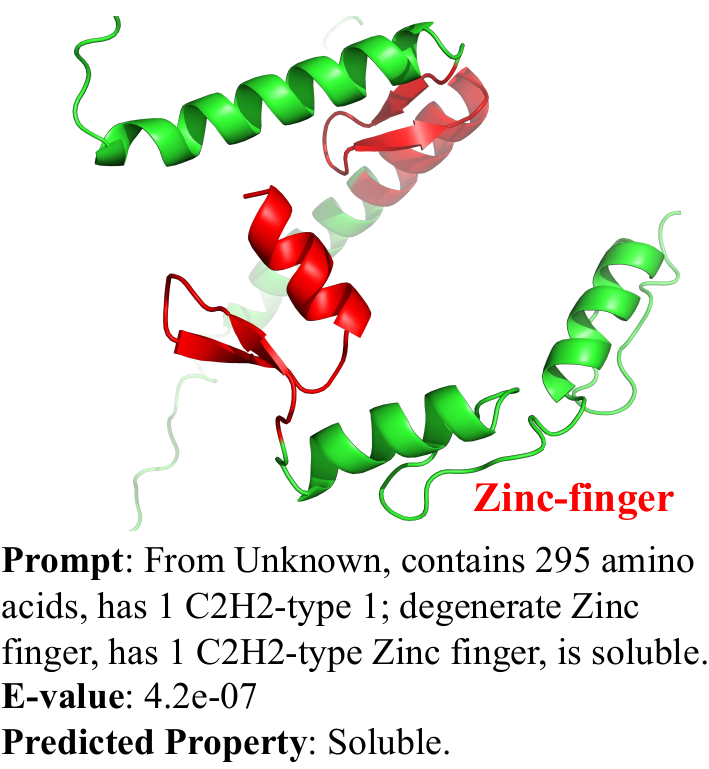}
    \includegraphics[width=0.23\textwidth]{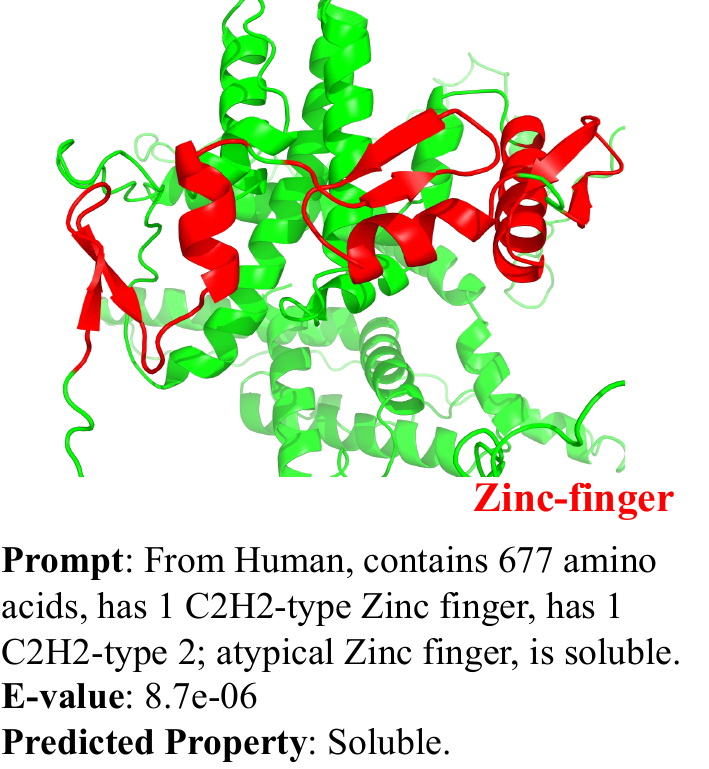}
    \vspace{-3ex}
    \caption{Visualization of jointly generation with domain and property annotations. {\method} is capable of integrating domain and property annotations.}
    \vspace{-3ex}
\label{fig:case_study_zinc_mambrane}
\end{figure*}
\subsection{Case study: Joint Generation with Domain and Property Annotations}
In this part, we explore the potential of {\method} in generating proteins guided by the flexible combination of different annotations. Specifically, we generate the zinc-finger proteins in membrane-bound and soluble by the model in Section~\ref{sec.prop_anno_gen} following the same protocol in Section~\ref{Protein Design with Domain Annotations}. Despite this challenging setting, the joint success rate $\text{SR}_{1}$ of generating proteins with zinc-finger and corrected property is $\text{SR}_{1} = 10.17\%$, However, given the current success rates of other models in generating zinc-finger proteins have approached zero, the task of producing such proteins with specific domain and property annotations remains unattainable by existing models. Zinc-finger domains are naturally occurring soluble proteins involved in DNA editing. Our generation of zinc-finger proteins anchored to the membrane underscores the efficacy of {\method} in producing novel, non-existent proteins. We showcase four examples of the generated results in Figure ~\ref{fig:case_study_zinc_mambrane}. As shown in Figure~\ref{fig:case_study_zinc_mambrane}, {\method} can successfully generate proteins guided by both domain and property annotations, demonstrating its potential in complex protein design tasks.

\subsection{Ablation Study}
\label{sec.abstudy}
To ascertain the contribution of each component towards the generation of the functional proteins, the ablation study reports the $\text{SR}_{1}$ of zinc-finger and immunoglobulin domains in the absence of each alignment loss, as presented in Table \ref{tab:ablation}. We observe that $\mathcal{L}_{ADC}$ is the key to the high SuccessRate of {\method}. The SuccessRate decreases to $0\%$ without $\mathcal{L}_{ADC}$, underscoring the importance of incorporating domain range into the learning framework. Furthermore, $\mathcal{L}_{APC}$ and $\mathcal{L}_{APM}$ also enhance the SuccessRate of generating high-quality immunoglobulin domains by $4.33\%$ and $3.33\%$, respectively. Moreover, $\mathcal{L}_{APC}$ and $\mathcal{L}_{APM}$ are more important to zinc-finger domain, by improving the $\text{SR}_{1}$ by  $12\%$ and $9.67\%$. %

\begin{table}[h]
  \centering
  \vspace{-2ex}
  \caption{Ablation study for each component. Performance of $\text{SR}_{1}$ for zinc-finger and immunoglobulin domains.}
  \vspace{-2ex}
  \resizebox{0.75\linewidth}{!}{
    \begin{tabular}{lllrr}
    \toprule
    $\mathcal{L}_{ADC}$ & $\mathcal{L}_{APC}$ & $\mathcal{L}_{APM}$ & \multicolumn{1}{l}{\textbf{Zinc-finger}} & \multicolumn{1}{l}{\textbf{Immunoglobulin}} \\
    \midrule
          & $\checkmark$ & $\checkmark$ & 0.00\%   & 0.33\% \\
    $\checkmark$ &       & $\checkmark$ &  39.00\% & 18.33\%  \\
    $\checkmark$ & $\checkmark$ &       & 41.33\% & 19.33\% \\
    \midrule
    $\checkmark$ & $\checkmark$ & $\checkmark$ & \textbf{51.00\%}  & \textbf{22.67\%} \\
    \bottomrule
    \end{tabular}%
    }
    \vspace{-3ex}
  \label{tab:ablation}%
\end{table}%

\section{Related Work}
\label{sec:Related Work}
\textbf{Moltimodal Representation Learning.} By harnessing the potential of extensive image-text pair data, the Contrastive Language-Image Pretraining (CLIP) model, as proposed by Radford et al.~\cite{radford21clip}, employs contrastive learning to align the representations between image and text modalities.Following by CLIP, many image-text pertaining model are proposed, such as BILP~\cite{li2022blip}, BLIP-2~\cite{li2023blip}, InstructBLIP~\cite{dai2023instructblip} and ClipCap~\cite{mokady2021clipcap}. Beyond the image-text pretraining, several studies introduce the more modalities, such as videos~\cite{hu2021videoclip}, audios~\cite{tang2023any} and even molecules~\cite{liu2022moleculestm, zhang2024atomas} into a unified representation. Specifically, for multimodal learning on protein sequences, OntoProtein~\cite{zhang2022ontoprotein} first learns protein representations by combining them with textual descriptions in a knowledge graph. 
ProtST~\cite{xu2023protst} constructs a large-scale dataset containing aligned pairs of protein sequences and property descriptions, and pretrain a protein-biotext model to improve performance on downstream predictive task and enables zero-shot retrieval.

\textbf{Protein Generation Model.}  With huge success of language models~\cite{devlin2019bert}, several studies~\cite{elnaggar2021protrans, shin2021protein, ferruz2022protgpt2,rives2019esm1}  treat protein sequences consisting chains of amino acids as a type of ``languages'' and pretrain models on millions of protein sequences. Upon the pretrained model, they generate the protein sequences in an autoregressive manner. In addition to the autoregressive model, Evodiff~\cite{Alamdari2023.09.11.556673} extracts the evolutionary information from protein sequences and proposes an 
evolution-guided diffusion model to generate protein sequences.
Given the significance of structural information for protein function, a group of methods~\cite{hsu2022learning, ingraham2019generative, tan2022generative, jing2020learning}, known as inverse folding, utilize the structure as a conditional input, allowing for the generation of amino acid sequences.
Some studies~\cite{shi2022protein, watson2023novo}  integrate both sequential and structural information and propose a co-design model that accepts sequences and structures as conditions. Recently, ProteinDT~\cite{liu2023proteindt} first proposes a text-sequence alignment framework, enabling its capabilities for text-guided protein generation and editing. Chroma~\cite{ingraham2022chroma} trains a protein caption model ProCap and utilize it as a classifier guidance to generate proteins via a diffusion model. 

Although ProteinDT~\cite{liu2023proteindt} and ProtST~\cite{xu2023protst} similarly align the protein and textual representations, the multi-level alignment framework enables PAAG to capture the both local and global properties of the protein. Furthermore, the momentum encoders and an additional matching loss assist PAAG to more effectively align the protein-level annotation with the proteins.

\vspace{-1ex}
\section{Conclusion}
\label{Conclusion}
This paper presents {\method}, a multi-modality framework that first incorporates the rich annotation information derived from protein database, achieving the superior performance in various applications, such as representation learning and annotation-guided protein design. Crucially, we demonstrate that it is possible to use the flexible combinations of various kinds of textual annotations to guide the protein design process. We hope that {\method} will expand the possibilities of protein design and establish a robust foundation for future advancements in protein-related applications. Future research directions include functional protein editing, co-design of protein sequence-structure within alignment frameworks, and exploring the potential of larger protein dataset with lower annotation quality.

\begin{acks}
The authors would like to thank Siying Xu (China Academy of Art) for her help in visualizing the framework and results. This work was jointly supported by the following projects: the National Natural Science Foundation of China (No. 62376276); Beijing Nova Program (No. 20230484278); Beijing Outstanding Young Scientist Program (No. BJJWZYJH012019100020098); the Fundamental Research Funds for the Central Universities, and the Research Funds of Renmin University of China (23XNKJ19). 
\end{acks}
\bibliographystyle{ACM-Reference-Format}
\bibliography{paag}

\appendix

\eat{\section{Related Work}
\label{appendix:related work}
\textbf{Difference compared with other text-protein multimodal representation learning:} 

The distinction between PAAG and ProtST lies in the incorporation of momentum contrastive (MoCo) \cite{he2020momentum} and an additional matching loss to facilitate the capture of fine-grained alignment between textual descriptions and protein sequences. MoCo establishes a dynamic dictionary utilizing a queue and a moving-average encoder, thereby enabling our model to learn better representations. While ProtST implements additional masked protein modeling loss and multimodal mask prediction loss to obtain residue-level properties, we demonstrate that {\method}-induced PLMs can achieve better results in downstream tasks. Additioanlly, our multi-level alignment framework can capture the functional parts in protein sequences.

ProteinDT can design protein sequences guided by the textual description. However, ProteinDT does not include the local information, which makes ProteinDT difficult to distinguish which part of protein is functional. The multi-level framework PAAG introduces explicitly aids PLMs in understanding the definition of a functional domain. Given that these functional domains directly exhibit specific functions, {\method} demonstrates superior performance in generating proteins based on the provided multi-level annotations.
}

\section{Experimental settings}
\label{appendix.experimental_settings}

\subsection{General settings}

\textbf{Backbone Models of {\method}:} We use ProtBert-BFD~\cite{elnaggar2021protrans} to initialize our protein encoder and SciBert~\cite{beltagy2019scibert} for the text encoder. Due to limited training data, we opt for a lighter decoder initialized by DistilProtBert~\cite{geffen2022distilprotbert}. Exploring a decoder with more parameters is a future direction.

\textbf{Training Configurations:} We train {\method} on {\dataset}, using the AdamW optimizer (with a learning rate of 3e-5 and zero weight decay) for 100 epochs. Our generation experiments are conducted on 16 NVIDIA Tesla A100-SX4-40GB GPUs.

\subsection{{\dataset} and template function $G(\cdot)$}
\label{appendix.dataset}
Table~\ref{tab:datastat} depicts additional statistics of {\dataset}. We demonstrate several example data samples in {\dataset} as well as the corresponding textual description generated by the template function $G(\cdot)$.
\begin{table}[htbp]
  \centering
  \vspace{-2ex}
  \caption{The statistics of {\dataset}}
  \vspace{-2ex}
  \resizebox{\linewidth}{!}{
\begin{tabular}{r|r|r|r}
\toprule
{\textbf{\# of proteins}} & {\textbf{\# of distinct domains}} & {\textbf{average \# of domains per protein}} & {\textbf{average length}} \\
\midrule
129,727    &    1,416        &      1.60      & 419.40 \\
\bottomrule
\end{tabular}%
}
\vspace{-3ex}
\label{tab:datastat}
\end{table}

The explanation of four property annotations:
\begin{itemize}[leftmargin=*]
    \item \textbf{protein\_name}: The protein name is a naming method used to describe the function, characteristics, or origin of a protein. These names usually contain information about the protein's structure, function, substrate specificity, and biological process.
    \item \textbf{organism\_name}: Organism names are used to identify and classify different species of living organisms, including bacteria, fungi, plants, and animals. These names usually consist of the genus and species of the organism, and sometimes include additional information such as strain or cultivar.
    \item \textbf{length}: The number of amino acids in the protein sequence.
    \item \textbf{SIMILARITY}: In the context of biology and protein classification, ``similarity'' refers to the shared characteristics or features among different proteins. This can include similar structures, functions, or evolutionary origins. Proteins with high similarity are often grouped into the same family or subfamily.
\end{itemize}

\begin{table*}[htbp]
  \begin{minipage}{0.6\textwidth}
  \centering
  \caption{Examples of annotations and description of protein\\ in ProtAnnotation}
  \vspace{-2ex}
  \resizebox{1.0\linewidth}{!}{
    \begin{tabular}{p{5em}|p{9.6em}|p{15em}|p{15em}}
    \toprule
    Entry name & \textcolor[rgb]{ .192,  .608,  .384}{Domain Annotation} & \textcolor[rgb]{ .871,  .235,  .212}{Property Annotation} & Textual Description G(D) \\
    \midrule
    Q8W4R8     & [38, 95], Ubiquitin-like; degenerate domain\newline{}[158, 459],PI3K/PI4K catalytic domain & \textbf{organism\_name}: Mouse-ear cress, \newline{}\textbf{protein\_name}: Phosphatidylinositol 4-kinase gamma 6, \newline{}\textbf{length}: 622, 
    \newline{}\textbf{SIMILARITY}: Belongs to the PI3/PI4-kinase family, Type II PI4K subfamily & This is \textcolor[rgb]{ .871,  .235,  .212}{Phosphatidylinositol 4-kinase gamma 6 protein}, from \textcolor[rgb]{ .871,  .235,  .212}{Mouse-ear cress}, belongs to \textcolor[rgb]{ .871,  .235,  .212}{the PI3/PI4-kinase family, Type II PI4K subfamily}, contains \textcolor[rgb]{ .871,  .235,  .212}{622 amino acids}, has \textcolor[rgb]{ .192,  .608,  .384}{1 Ubiquitin-like; degenerate domain}, has \textcolor[rgb]{ .192,  .608,  .384}{1 PI3K/PI4K catalytic domain}. \\
    \midrule
    Q8XAW7     & [5, 241], ABC transporter domain\newline{}[252, 495], ABC transporter domain & \textbf{organism\_name}: Escherichia coli O157:H7, \newline{}\textbf{protein\_name}: Ribose import ATP-binding protein RbsA 1, \newline{}\textbf{length}: 501, \newline{}\textbf{SIMILARITY}: Belongs to the ABC transporter superfamily, Ribose importer (TC 3,A,1,2,1) family & This is \textcolor[rgb]{ .871,  .235,  .212}{Ribose import ATP-binding protein RbsA 1 protein}, from \textcolor[rgb]{ .871,  .235,  .212}{Escherichia coli O157:H7}, belongs to \textcolor[rgb]{ .871,  .235,  .212}{the ABC transporter superfamily, Ribose importer (TC 3,A,1,2,1) family}, contains \textcolor[rgb]{ .871,  .235,  .212}{501 amino acids}, has \textcolor[rgb]{ .192,  .608,  .384}{2 ABC transporter domain}. \\
    \eat{
    \midrule
    Q8Y2D7     & [254, 288], FPG-type Zinc finger & \textbf{organism\_name}: Pseudomonas solanacearum, \newline{}\textbf{protein\_name}: Formamidopyrimidine-DNA glycosylase, \newline{}\textbf{length}: 288, \newline{}\textbf{SIMILARITY}: Belongs to the FPG family & This is \textcolor[rgb]{ .871,  .235,  .212}{Formamidopyrimidine-DNA glycosylase protein}, from \textcolor[rgb]{ .871,  .235,  .212}{Pseudomonas solanacearum}, belongs to \textcolor[rgb]{ .871,  .235,  .212}{the FPG family}, contains \textcolor[rgb]{ .871,  .235,  .212}{288 amino acids}, has \textcolor[rgb]{ .192,  .608,  .384}{1 FPG-type Zinc finger}. \\}
    \bottomrule
    \end{tabular}%
    \label{tab:dataexample}%
    }
    \end{minipage}
    \vspace{-3ex}
    \begin{minipage}{0.33\textwidth}
      \centering
      \begin{minipage}[t]{\linewidth}
        \caption{The success rate on the prompt with single domain.}
        \vspace{-2ex}
        \resizebox{0.85\columnwidth}{!}{\begin{tabular}{l|r|r}
        \toprule
        \textbf{Method} & \multicolumn{1}{l|}{\textbf{SR$_{1}$}(Zinc finger)} & \multicolumn{1}{l}{\textbf{SR$_{1}$}(Immunoglobulin)} \\
        \midrule
        \textbf{Chroma} & 0.33\% & 0\% \\
        \midrule
        \textbf{ProteinDT} & 1.67\% & 22\% \\
        \midrule
        \textbf{PAAG}  & 22.67\% & 51\% \\
        \midrule
        \textbf{PAAG}$_{\text{single domain}}$ & 18.33\% & 46\% \\
        \bottomrule
        \end{tabular}}
        \label{tab:singledomain}%
      \end{minipage}
      \vspace{1ex}
      \centering
      \begin{minipage}[t]{\linewidth}
        \caption{Results of the proteins unconditionally generated by fine-tuned {\method}.}
        \vspace{-2ex}
        \resizebox{0.85\columnwidth}{!}{
        \begin{tabular}{l|l|l|r}
        \toprule
        \textbf{Model} & \textbf{Distinc-2} & \textbf{Diversity} & \multicolumn{1}{l}{\textbf{Novelty}} \\
        \midrule
        \textbf{Natural} & 0.4309(0) & 0.829(0) & \multicolumn{1}{l}{-} \\
        \midrule
        \textbf{ProGen} & 0.3003(30.31\%) & 0.845(1.93\%) & 0.374 \\
        \midrule
        \textbf{Chroma} & 0.3211(25.48\%) & 0.855(3.14\%) & 0.638 \\
        \midrule
        \textbf{PAAG}$_{\text{before ft}}$ & 0.4314(0.12\%) & 0.815(1.69\%) & 0.766 \\
        \midrule
        \textbf{PAAG}$_{\text{after ft}}$ & 0.4524(4.99\%) & 0.822(0.84\%) & 0.674 \\
        \bottomrule
        \end{tabular}}%
        \label{tab:finetuneuncondition}
      \end{minipage}
    \end{minipage}
  \vspace{-0.5ex}
\end{table*}%

\subsection{Biological background of zinc-finger and immunoglobulin domain}
We here introduce zinc-finger and immunoglobulin domains, highlighting their biological significance and functions in biology.

\begin{itemize}[leftmargin=*]
    \item \textbf{Zinc-finger:} Zinc finger~\cite{klug1987zinc} domains are compact protein motifs known for binding DNA, RNA, proteins, and lipids, with binding characteristics influenced by amino acid sequence, linker structure, and number of fingers. Despite appearing in various protein families, they maintain stable structures and are crucial in processes like gene transcription and mRNA trafficking.

    \item \textbf{Immunoglobulin domain: }Immunoglobulin (Ig) domains play a key role in protein-protein interactions, especially within the immune system, where they help recognize, bind, and neutralize antigens. These domains have a conserved structure found in a variety of proteins, including antibodies and cell adhesion molecules.
\end{itemize}

\begin{figure}
    \centering
    \includegraphics[width=0.2\textwidth]{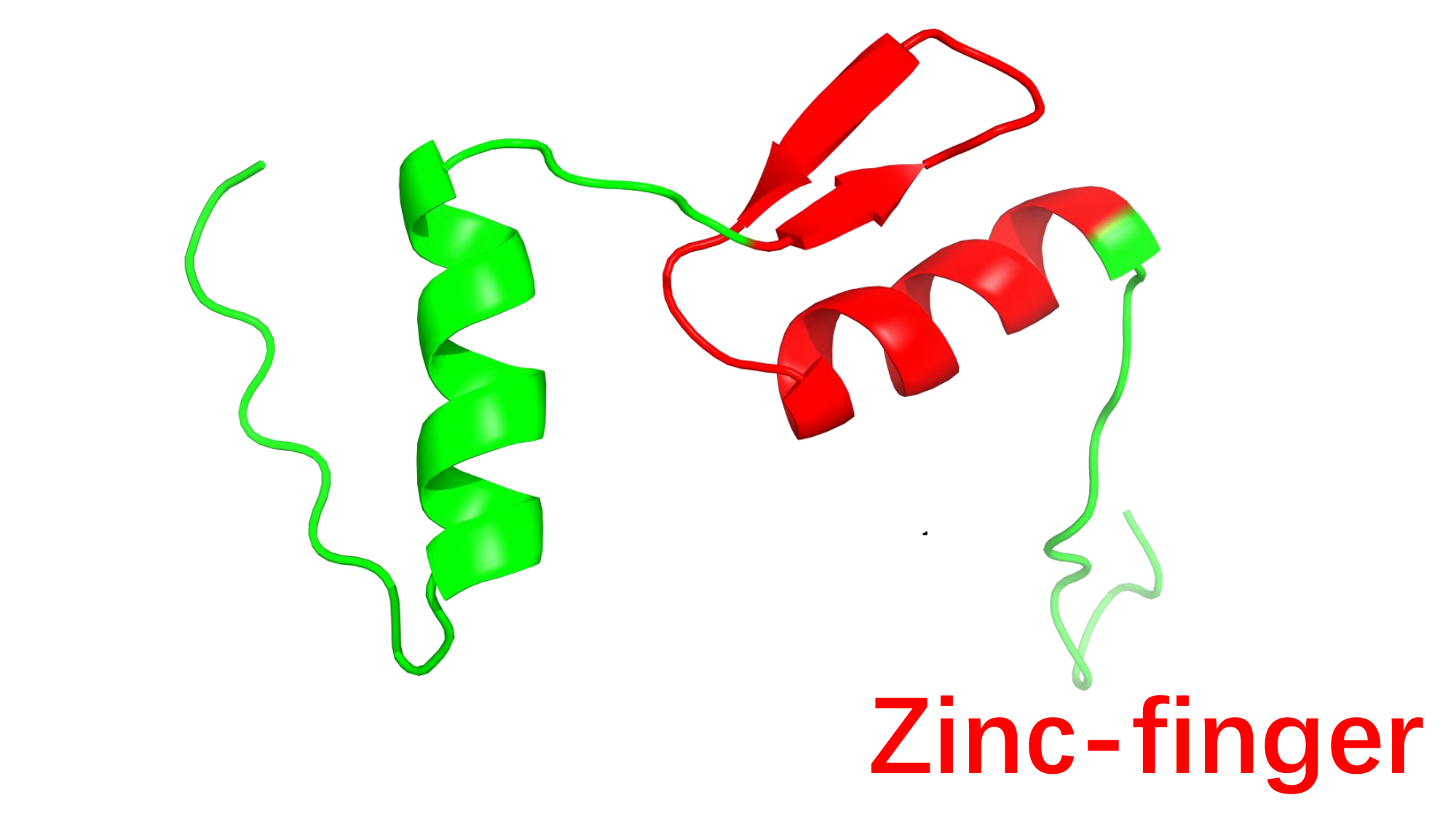}
    \includegraphics[width=0.2\textwidth]{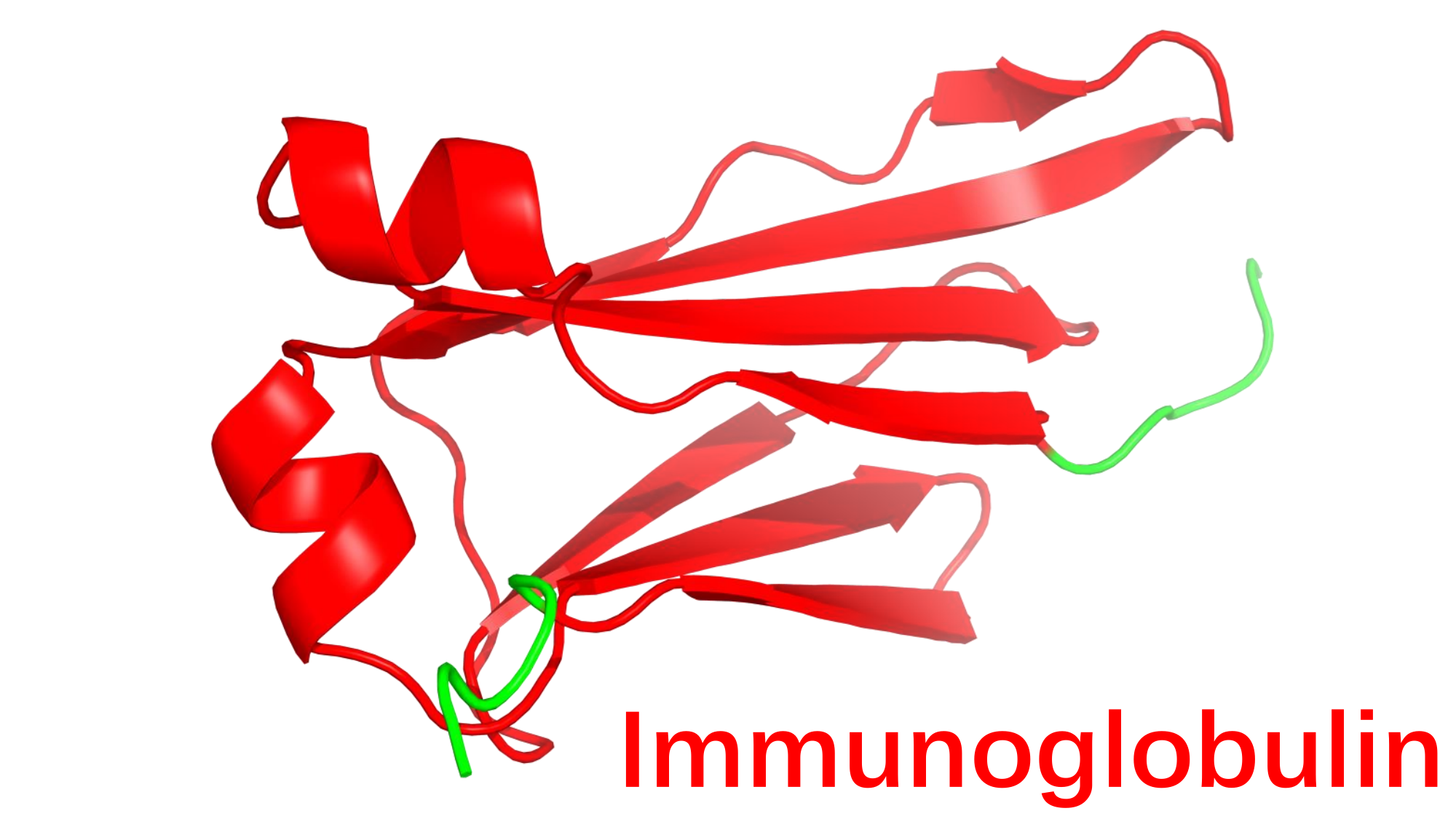}
        \vspace{-3ex}
    \caption{Visualization of typical proteins containing Zinc-finger (P26634) and Immunoglobulin (A0M8Q6).}
    \vspace{-4ex}
\label{fig:Biological_background_example_zinc_ig}
\end{figure}

Examples of Zinc-finger and Ig domains are shown in Figure~\ref{fig:Biological_background_example_zinc_ig}.

\subsection{The evaluation metric for the unconditional generation}
\label{appendix.evaluationmetric}

In the computation of the Distinct-n metric, we assign a value of $n$ to $2$. For the Novelty metric, the default parameters of Mmseq2 are employed with an exception for the e-value threshold, which is designated as $100$. In the case of the Diversity metric, the default parameters of Mmseq2 are again utilized, but with alterations to the e-value threshold and sensitivity, set to $1000000$ and $15$ respectively. During the application of Mmseq2 for the calculation of Novelty and Diversity metrics, the dissimilarity for unmatched sequences is assumed to be $1.0$.

\subsection{More details of baselines}
\label{appendix.baselines}
\eat{ProGen~\cite{madani2020progen} first forms each property of protein as keyword tags. In the training stage, ProGen prepends the keyword tags to the corresponding amino acid sequence, and try to reconstruct the amino acid sequence in the auto-regressive manner. ProGen consequently has the ability to generate the protein sequence given the keyword tags. During controllable generation, ProGen first translates desired properties into keyword tags, then can controllably generate protein sequence.
}
ProGen~\cite{madani2020progen} represents each protein property as keyword tags, which are prepended to amino acid sequences during training to enable auto-regressive reconstruction. This allows ProGen to generate protein sequences from keyword tags. For controllable generation, ProGen translates desired properties into keyword tags to guide sequence generation.

\eat{Chroma~\cite{ingraham2022chroma} is also a protein design model that but focus more on generating the protein backbone structure. Specifically, Chroma is a diffusion model that can generate protein backbone with desired properties and functions by accepting the classifier guidance. Therefore, Chroma introduces ProCap, a protein-to-text model, as the classifier guidance to enable text-guided protein design. With an additional sequence sampling model, Chroma is capable of sampling both structure and sequence based on the conditions.
}
Chroma~\cite{ingraham2022chroma} is a protein design model focusing on generating protein backbones. It uses classifier guidance and ProCap, a protein-to-text model, for text-guided designs. With a sequence sampling model, Chroma can generate both structure and sequence based on specified conditions.

ProteinDT~\cite{liu2023proteindt} is a text-guided protein design model that aligns protein and text representations, then uses aligned text representations to design protein sequences via autoregressive and diffusion models. For fair comparison, we use autoregressive ProteinDT as our baseline. In unconditional generation, due to the lack of a specific template in ProteinDT, we randomly sample prompts from its dataset to evaluate the quality of generated proteins. In conditional experiments, ProteinDT receives the same prompts as PAAG.

In unconditional generation, {\method} accepts only the protein sequence length $l$ and the order $k$ of protein generated, using the template: ''protein number k, contains l amino acids.'' For generating proteins with functional domains, we randomly sample organism names, lengths, and similarities from natural proteins with the corresponding domains. Generative hyper-parameters are specified in \ref{appendix.training_configurations}. Using property annotations, we apply the template function $G()$ to describe test data from the Deeploc~\cite{almagro2017deeploc} dataset, enabling {\method} to generate proteins with specified properties.

\eat{In unconditional generation, {\method} only accept the length $l$ of the protein sequence. We also define this is the $k$-th protein we generated Consequently, we set the template function as ``protein number k, contains l amino acids.''

While generate proteins with functional domains, we randomly sample the organism name, length and similarity from natural proteins that have corresponding functional domain. We specify the generative hyper-parameters in \ref{appendix.training_configurations}. When given property annotations. Deeploc~\cite{almagro2017deeploc} contains  training data and test data in its original split. We utilize template function $G()$ to form the textual descriptions of test data. Then {\method} can generate protein with properties given these descriptions.
}

\eat{
\subsection{Dataset Statistics}
\label{appendix: dataset stats}
We further introduce the class distribution of the dataset we used as property prediction tasks. Since we only conduct classification problem on localization prediction tasks, here we will show the class distribution of these two datasets in Table \ref{tab:bin_statistics} and \ref{tab:sub_statistics}. 

\begin{table}[htbp]
  \centering
  \caption{Class distribution of binary location prediction dataset.}
    \begin{tabular}{l|r}
    \toprule
    \textcolor[rgb]{ .2,  .2,  .2}{Location} & \multicolumn{1}{l}{Number of proteins} \\
    \midrule
    Soluble & 5109 \\
    \midrule
    Membrane-bound & 3553 \\
    \bottomrule
    \end{tabular}%
  \label{tab:bin_statistics}%
\end{table}%

}

\subsection{Training configurations}
\label{appendix.training_configurations}
\subsubsection{Predictive Task}

\eat{In downstream prediction tasks, {\method} utilizes the embedding in the aligned space. Specifically, we not only use the protein encoder to extract the protein representations. We also keep the projector head to extract the more informative aligned features. The hyperparameters for the predictive task concerning PAAG-ProtBert and PAAG-ESM-2 can be found in Table ~\ref{tab:predictive_protbert_hyperparam} and Table ~\ref{tab:predictive_esm-2_hyperparam}, respectively.
}
In downstream prediction tasks, {\method} uses embeddings in the aligned space by employing both the protein encoder and the projector head for more informative aligned features. Hyperparameters for predictive tasks using PAAG-ProtBert and PAAG-ESM-2 are detailed in Table ~\ref{tab:predictive_protbert_hyperparam} and Table ~\ref{tab:predictive_esm-2_hyperparam}.

\subsubsection{Protein Design with Domain Annotations}
\label{appendix:proteindesign}
While training {\method} with ProtAnnotation, we incorporate momentum contrast~\cite{he2020momentum} with a queue size of 16,384 and momentum of 0.995. The learnable temperature in contrastive learning is set to 0.07, and our latent space is aligned to 256 dimensions with a linear layer. Following \cite{li2022blip}, we initialize learnable alpha at 0.4 for soft labeling. When fine-tuning on datasets with only zinc-finger and immunoglobulin domains, we maintain these hyperparameters but reduce the queue size to 4,096 due to less training data. Additionally, weight decay for AdamW is set at 0.05.

In our generative task, we use nucleus sampling in the decoder, sampling amino acids based on probability rather than always selecting the highest probability ones. The Top-p hyperparameter is set to 0.9, meaning the decoder considers the top 90\% probability mass. We also set the decoder's repetition penalty to 1.2.

\subsubsection{Protein Design with Property Annotations}
\label{appendix:protannotation config}
Due to limited training data, we fine-tune the models from Section \ref{Protein Design with Domain Annotations} for 100 epochs. We retain the ProtAnnotation training hyperparameters but lower the learning rate to 1e-5 without warm-up.

For binary localization properties, soluble and membrane-bound, we set template function $G()$ as ``is soluble'' or ``is membrane-bound''.

\begin{table}[htbp]
    \centering
    \caption{Configurations of full-model tuning of PAAG-ProtBert. \emph{Abbr.}, lr.: learning rate; lrr.: learning rate ratio; dr.: dropout rate; wd.: weight decay; bs.: batch size; MSE: mean squared error; CE: cross entropy; BCE: binary cross entropy.}
    \vspace{-2ex}
        \resizebox{0.8\linewidth}{!}{\begin{tabular}{l|cccccccc}
            \toprule
            \bf{Task}  & \bf{lr.} & \bf{lrr.} & \bf{dr.} & \bf{wd.} & \bf{bs.} & \bf{\#epochs} & \bf{loss} \\
            \midrule
            \multicolumn{8}{c}{\bf{Localization}} \\
            \midrule
            \bf{Bin} & $2.0 \times 10^{-5}$ & 0.15 & 0 & 0 & 4 & 100 & BCE \\
            \bf{Sub} & $2.0 \times 10^{-5}$ & 0.15 & 0.1 & 0 & 4 & 100 & CE \\
            \midrule
            \multicolumn{8}{c}{\bf{Fitness}} \\
            \midrule
            \bf{$\rho$-lac} & $2.0 \times 10^{-4}$ & 0.02 & 0 & $3.0 \times 10^{-4}$ & 36 & 100 & MSE \\
            \bf{AAV} & $9.0 \times 10^{-4}$ & 0.02 & 0 & 0 & 36 & 100 & MSE \\
            \bf{Thermo} & $2.0 \times 10^{-4}$ & 0.02 & 0 & 0 & 6 & 100 & MSE \\
            \bf{Flu} & $2.0 \times 10^{-4}$ & 0.02 & 0.3 & $4.0 \times 10^{-4}$ & 36 & 100 & MSE \\
            \bf{Sta} & $5.0 \times 10^{-4}$ & 0.02 & 0.7 & 0 & 36 & 100 & MSE \\
            \bottomrule
        \end{tabular}}
        \vspace{-3ex}
        \label{tab:predictive_protbert_hyperparam}
\end{table}

\eat{
\begin{table*}[htbp]
    \centering
    \caption{Configurations of full-model tuning of PAAG-ESM-1b on three task types. \emph{Abbr.}, lr.: learning rate; lrr.: learning rate ratio; dr.: dropout rate; wd.: weight decay; bs.: batch size; MSE: mean squared error; CE: cross entropy; BCE: binary cross entropy.}
        \begin{tabular}{l|cccccccc}
            \toprule
            \bf{Task} & \bf{lr.} & \bf{lrr.} & \bf{dr.} & \bf{wd.} & \bf{bs.} & \bf{\#epochs} & \bf{loss} \\
            \midrule
            \multicolumn{8}{c}{\bf{Localization}} \\
            \midrule
            \bf{Bin} & $4.0 \times 10^{-4}$ & 0.1 & 0 & 0 & 4 & 100 & BCE \\
            \bf{Sub} & $2.0 \times 10^{-4}$ & 0.1 & 0 & 0 & 4 & 100 & CE \\
            \midrule
            \multicolumn{8}{c}{\bf{Fitness}} \\
            \midrule
            \bf{$\rho$-lac} & $6.0 \times 10^{-5}$ & 0.1 & 0 & 0 & 16 & 100 & MSE \\
            \bf{AAV} & $2.0 \times 10^{-4}$ & 0.1 & 0 & 0 & 16 & 100 & MSE \\
            \bf{Thermo} & $2.0 \times 10^{-4}$ & 0.1 & 0 & 0 & 2 & 100 & MSE \\
            \bf{Flu} & $4.0 \times 10^{-4}$ & 0.1 & 0 & 0 & 16 & 100 & MSE \\
            \bf{Sta} & $2.0 \times 10^{-5}$ & 0.1 & 0 & 0 & 16 & 100 & MSE \\
            \bottomrule
        \end{tabular}
        \label{tab:predictive_esm-1b_hyperparam}
\end{table*}
}
\begin{table}[htbp]
    \centering
    \caption{Hyperparameters for PAAG-ESM-2 in downstream tasks. \eat{\emph{Abbr.}, lr.: learning rate; lrr.: learning rate ratio; dr.: dropout rate; wd.: weight decay; bs.: batch size; MSE: mean squared error; CE: cross entropy; BCE: binary cross entropy.}}
    \vspace{-2ex}
        \resizebox{0.8\linewidth}{!}{\begin{tabular}{l|cccccccc}
            \toprule
            \bf{Task} & \bf{lr.} & \bf{lrr.} & \bf{dr.} & \bf{wd.} & \bf{bs.} & \bf{\#epochs} & \bf{loss} \\
            \midrule
            \multicolumn{8}{c}{\bf{Localization}} \\
            \midrule
            \bf{Bin} & $1.0 \times 10^{-5}$ & 0.15 & 0 & 0 & 4 & 100 & BCE \\
            \bf{Sub} & $2.0 \times 10^{-5}$ & 0.15 & 0 & 0 & 4 & 100 & CE \\
            \midrule
            \multicolumn{8}{c}{\bf{Fitness}} \\
            \midrule
            \bf{$\rho$-lac} & $1.0 \times 10^{-4}$ & 0.02 & 0 & 0 & 16 & 100 & MSE \\
            \bf{AAV} & $8.0 \times 10^{-4}$ & 0.02 & 0 & 0 & 16 & 100 & MSE \\
            \bf{Thermo} & $3.0 \times 10^{-4}$ & 0.02 & 0.6 & 0 & 2 & 100 & MSE \\
            \bf{Flu} & $3.0 \times 10^{-4}$ & 0.02 & 0 & 0 & 16 & 100 & MSE \\
            \bf{Sta} & $8.0 \times 10^{-5}$ & 0.02 & 0 & 0 & 16 & 100 & MSE \\
            \bottomrule
        \end{tabular}}
        \vspace{-3ex}
        \label{tab:predictive_esm-2_hyperparam}
\end{table}

\subsection{Additional experimental results}

\eat{
\subsubsection{Results of relevant improvement in property prediction tasks}
\label{sec: relevant improvement }
The relative improvement of backbone without alignment is shown in Table \ref{fig:prediction}.

\begin{figure}
\centering
\includegraphics[width=0.85\linewidth]{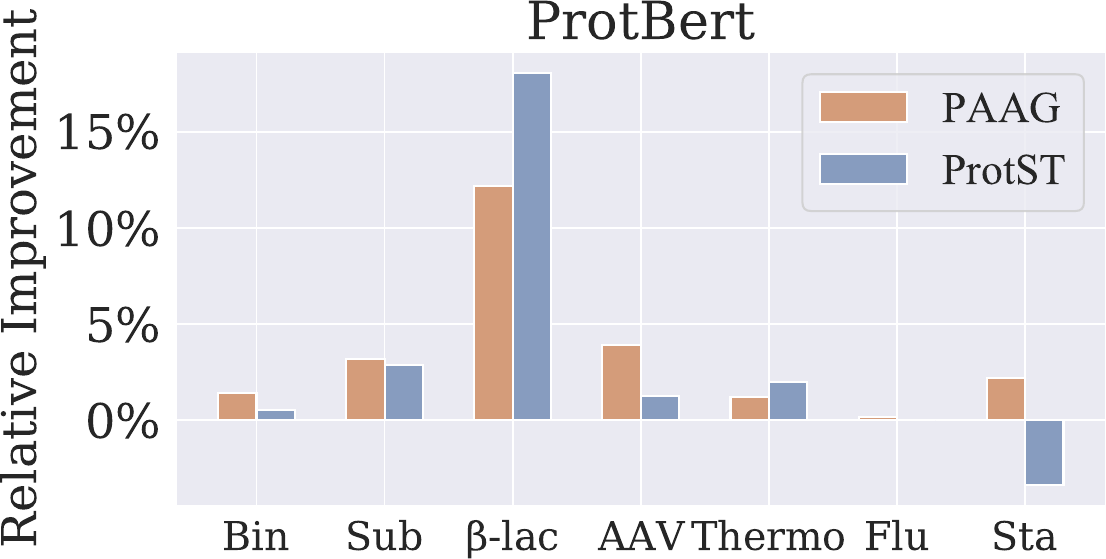}
\vspace{3ex}
\includegraphics[width=0.85\linewidth]{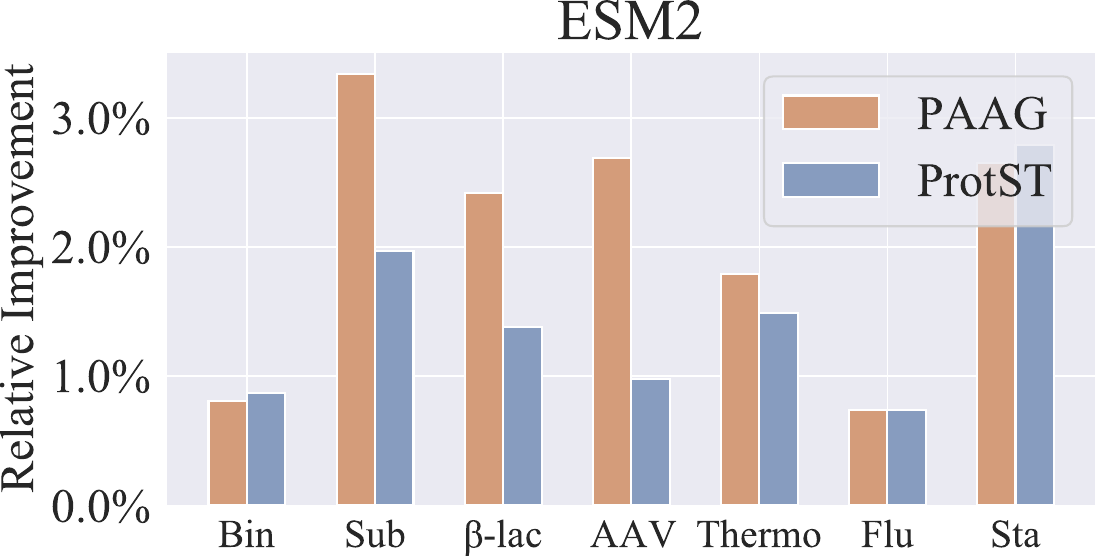}
\caption{Relevant improvement of {\method} and ProtST compared with vanilla pretrained protein encoders. ProtST is observed a negative transfer in Sta task when using ProtBert, while {\method} exhibits improved performance across all tasks on both encoders. }
\label{fig:prediction}
\vspace{-3.5ex}
\end{figure}
}

\subsubsection{Results of generating proteins with EGF-like domain}
We here generates $300$ functional proteins with EGF-like domain for ProGen, ProteinDT and PAAG. The results are in Table~\ref{tab:egf}.
\label{appendix.egf}
\begin{table}[htbp]
  \centering
  \caption{Success Rate of generating proteins with EGF-like domain.}
  \vspace{-2ex}
    \resizebox{0.9\columnwidth}{!}
    {\begin{tabular}{lrrrrrr}
    \toprule
    \textbf{EFG-like domain} & \multicolumn{1}{l}{\textbf{SR$_{100}$}} & \multicolumn{1}{l}{\textbf{SR$_{10}$}} & \multicolumn{1}{l}{\textbf{SR$_{1}$}} & \multicolumn{1}{l}{\textbf{SR$_{0.1}$}} & \multicolumn{1}{l}{\textbf{SR$_{0.01}$}} & \multicolumn{1}{l}{\textbf{SR$_{0.001}$}} \\
    \midrule
    \textbf{ProGen} & 1\%   & 1\%   & 1\%   & 1\%   & 1\%   & 1\% \\
    \textbf{ProteinDT} & 0\%   & 0\%   & 0\%   & 0\%   & 0\%   & 0\% \\
    \textbf{PAAG}  & \textbf{28.67\%} & \textbf{28.33\%} & \textbf{26.00\%} & \textbf{22.33\%} & \textbf{17.33\%} & \textbf{11.67\%} \\
    \bottomrule
    \end{tabular}%
    }
    \vspace{-3ex}
  \label{tab:egf}%
\end{table}%

\subsubsection{More evaluation on the proteins generated by fine-tuned {\method}}
\label{appendix.fine-tuned unconditional}
To ensure fine-tuned PAAG isn't just memorizing the training set, we evaluate distinct-2, diversity, and novelty for unconditionally generated sequences. Table \ref{tab:finetuneuncondition} shows that, even after fine-tuning, PAAG produces proteins with higher novelty than Chroma and ProGen, indicating it doesn’t overfit, thanks to label smoothing. Additionally, fine-tuned PAAG's diversity is closest to natural proteins, suggesting it captures similar amino acid distributions.

\end{document}